\documentstyle[pra,aps,twocolumn]{revtex}
                      
\newcommand{\ket}[1]{\left|#1\right>} 
\newcommand{\bra}[1]{\left<#1\right|} 
 
\newcommand{\f}[1]{\mbox{\boldmath$#1$}}
\newcommand{\fk}[1]{\mbox{\boldmath$\scriptstyle#1$}}

\begin{document} 
\title{Quantum radiation at finite temperature}
\author{Ralf Sch\"utzhold, G\"unter Plunien, and Gerhard Soff}
\address{Institut f\"ur Theoretische Physik, Technische  Universit\"at
Dresden, D-01062  Dresden, Germany\\
Electronic address: {\tt schuetz@theory.phy.tu-dresden.de}}
\date{\today}
\maketitle
\begin{abstract}
We investigate the phenomenon of quantum radiation -- i.e.~the
conversion of (virtual) quantum fluctuations into (real) particles
induced by dynamical external conditions -- for an initial thermal
equilibrium state. 
For a resonantly vibrating cavity a rather strong
enhancement of the number of generated particles 
(the dynamical Casimir effect) 
at finite temperatures is observed.
Furthermore we derive the temperature corrections to the energy
radiated by a single moving mirror and an oscillating bubble within a
dielectric medium as well as the number of created particles within the
Friedmann-Robertson-Walker universe.
Possible implications and the relevance for experimental tests are
addressed. 
\end{abstract} 

%%%%%%%%%%%%%%%%%%%%%%%%%%%%%%%%%%%%%%%%%%%%%%%%%%%%%%%%%%%%%%%%%%%%%%%%%%%%%%
%%%%%%%%%%%%%%%%%%%%%%%%%%%%%%%%%%%%%%%%%%%%%%%%%%%%%%%%%%%%%%%%%%%%%%%%%%%%%%%
\section{Introduction}\label{intro}
%%%%%%%%%%%%%%%%%%%%%%%%%%%%%%%%%%%%%%%%%%%%%%%%%%%%%%%%%%%%%%%%%%%%%%%%%%%%%%%
%%%%%%%%%%%%%%%%%%%%%%%%%%%%%%%%%%%%%%%%%%%%%%%%%%%%%%%%%%%%%%%%%%%%%%%%%%%%%%%

Motivated by previously obtained results \cite{letter} we generalize
the canonical formalism adopted there towards the application to
further scenarios and establish the derivations in more detail: 

One of the main consequences of quantum theory is the existence of a  
non-trivial vacuum state. In contrast to the classical theory the
quantum fields undergo fluctuations even in their state of lowest
energy (the ground state) -- the so-called vacuum fluctuations.
These fluctuations have measurable consequences:
E.g., if the fields are constraint by the presence of external
conditions, the energy associated to these fluctuations (the
zero-point energy) may change owing to the imposed external
conditions. 
As a result the quantum field may exert a force onto the external
conditions in order to minimize its energy.   
The most prominent example for such a force is the Casimir
\cite{casimir} 
effect which predicts the attraction of two parallel perfectly
conducting and neutral plates (i.e.~mirrors) placed in the vacuum of  
the electromagnetic field.  
The prediction of this striking effect has been verified
experimentally with relatively high accuracy
\cite{lamoreaux,mohideen}.

A different -- not less interesting -- effect has not yet been
rigorously verified in an experiment:
The impact of the external conditions may also induce a conversion of
the virtual quantum fluctuations of the field into real particles
-- the phenomenon of quantum radiation.
As examples for such external conditions giving rise to the creation
of particles we may consider moving mirrors, time-dependent
dielectrics or gravitational fields.

Various investigations have been devoted to this topic during the last 
decades, in order to mention only some of the most important initial
papers in chronological order:
In 1970 Moore \cite{moore} presented the first explicit calculation
of the quantum radiation on the basis of two 1+1 dimensional moving
mirrors. In this pioneering work he exploited the conformal
invariance of the scalar field in 1+1 dimensions.
Based on this result Fulling and Davies 
\cite{fulling+davies} presented a calculation of the
radiation of a single moving mirror (again in 1+1 dimensions) and
pointed out the close analogy to the Hawking radiation.
In 1982 Ford and Vilenkin \cite{ford+vilenkin} succeeded to develop a
method for the calculation of the radiation generated by a moving
mirror in higher dimensions, i.e., without exploiting the conformal
invariance.

For the experimental verification of the phenomenon of quantum
radiation the scenario of a closed cavity might be most promising
since one may exploit a resonance enhancement in this situation.
The particle production inside a resonantly vibrating cavity has
already been considered by several authors, 
see
e.g.~\cite{dodonov,law,trembling,kardar,jaekel,ji,yang,leung,geklaut,diplom} 
as well as \cite{milonni,bordag1,bordag2} for reviews. 

However, most of the investigations of quantum radiation are
restricted to the vacuum state, i.e.~to zero temperature.
But in view of an experimental verification it is essential to study
the finite temperature effects.  
Realistic calculations of thermal effects on quantum radiation 
within the framework of quantum field theory of time-dependent systems 
at finite temperature are not yet available.    

The remedy of this deficiency is the main intention of the present
article: 
In Sec.~\ref{formalism} we set up the basic formalism for the quantum
treatment of external conditions at finite temperatures.
The developed methods are applied in Section \ref{trembling} to the
scenario of a trembling cavity. In Sec.~\ref{resonantly} we focus on
the resonance case and derive the number of created particles. 
Another scenario giving rise to quantum radiation -- a dynamical
dielectric medium -- is considered in Section \ref{dielectric}. 
In Sec.~\ref{robertson} we demonstrate the flexibility of the
canonical approach presented in Section \ref{formalism} by calculating
the finite temperature corrections to the particle production in yet
another example scenario -- the Friedmann-Robertson-Walker universe.
We shall close with a summary, some conclusions, a discussion, and an
outlook. 

Throughout this article natural units with
\begin{eqnarray}
\hbar=c=G_{\rm N}=k_{\rm B}=\epsilon_0=\mu_0=1
\end{eqnarray}
will be used. The signature of the Minkowski metric is chosen
according to $g_{\mu\nu}={\rm diag}(+1,-1,-1,-1)$.

%%%%%%%%%%%%%%%%%%%%%%%%%%%%%%%%%%%%%%%%%%%%%%%%%%%%%%%%%%%%%%%%%%%%%%%%%%%%%%%
%%%%%%%%%%%%%%%%%%%%%%%%%%%%%%%%%%%%%%%%%%%%%%%%%%%%%%%%%%%%%%%%%%%%%%%%%%%%%%%
\section{General formalism}\label{formalism}
%%%%%%%%%%%%%%%%%%%%%%%%%%%%%%%%%%%%%%%%%%%%%%%%%%%%%%%%%%%%%%%%%%%%%%%%%%%%%%%
%%%%%%%%%%%%%%%%%%%%%%%%%%%%%%%%%%%%%%%%%%%%%%%%%%%%%%%%%%%%%%%%%%%%%%%%%%%%%%%

The objective is to investigate quantized bosonic fields obeying
linear equations of motion under the influence of external conditions.      
At asymptotic times $|t|\uparrow\infty$ the external conditions,
which are treated classically (not quantized), are assumed to approach
a static $\hat H_0$-configuration, where the asymptotic Hamiltonian
$\hat H_0$ can be diagonalized via a suitable particle definition. 
Initially the state of the quantum system is supposed to be described
by  thermal equilibrium at a given temperature $T$, which might be
realized through the coupling to a corresponding heat bath. 
However, the coupling to the heat bath has to be switched off before
the external conditions undergo dynamical changes in order to avoid
relaxation processes (closed system).
In general the time-dependence of the external conditions cause  
the state of the quantum system to leave the initial thermal
equilibrium.
Accordingly, the calculation of the expectation value of relevant
observables, e.g.~the number of particles, before and after the
dynamics may deviate.    
These differences can be interpreted as particles that are created
or even annihilated by the dynamical external conditions.

%%%%%%%%%%%%%%%%%%%%%%%%%%%%%%%%%%%%%%%%%%%%%%%%%%%%%%%%%%%%%%%%%%%%%%%%%%%%%%%
\subsection{Interaction picture}\label{interaction}
%%%%%%%%%%%%%%%%%%%%%%%%%%%%%%%%%%%%%%%%%%%%%%%%%%%%%%%%%%%%%%%%%%%%%%%%%%%%%%%

The following calculations are most suitably performed in the
interaction representation. Accordingly, the dynamics of all operators 
$\hat X$ corresponding to observables are governed by the undisturbed 
Hamiltonian $\hat H_0$
\begin{eqnarray}
\frac{d\hat X}{dt}=i\left[\hat H_0,\hat X \right]
+\left(\frac{\partial\hat X}{\partial t}\right)_{\rm exp}
\,.
\end{eqnarray}
This Hamiltonian $\hat H_0$ describes the complete dynamics of the system
at asymptotic times and can be diagonalized via a suitable particle
definition
\begin{eqnarray}
\hat H (|t|\uparrow\infty)
=
\hat H_0
=
\sum\limits_I\omega_I\,\hat N_I+E_0
=
\omega_I\,\hat N_I+E_0
\,,
\end{eqnarray}
where $E_0$ denotes the (divergent) zero-point energy.
In most of the following formulas we make use of a
generalized sum convention and drop the summation signs
by declaring that one has to sum over all indices that do 
not occur at both sides of the equation.
Equations with the same index appearing at both sides are valid for
all possible values of this index.
 
The index $I$ contains a complete set 
of quantum numbers labeling the different particle modes, e.g.
$I=\{\f{k}\}$ or $I=\{\omega,\ell,m\}$ etc.
The particle energies are given by $\omega_I$.  
For a thermodynamical consideration we have to describe the 
state of the field by the statistical operator $\hat\varrho$.
In the interaction picture the time evolution of this density matrix 
is given by the von Neumann equation
\begin{eqnarray}
\label{neumann}
\frac{d\hat\varrho}{dt}=-i\left[\hat H_1,\hat\varrho\right]
\, .
\end{eqnarray}
The perturbation Hamiltonian $\hat H_1$ governs the influence of the
variation of the external conditions upon the quantized field.
Note, that this equation describes the time evolution of a closed
quantum system, i.e.~no measurements (etc.) take place during the dynamics. 
It leads to an unitary time evolution operator and therefore does also not
contain relaxation processes, etc.
Measurements and relaxations would change the probabilities $w_A$
of the statistical operator 
\begin{eqnarray}
\hat\varrho(t)=w_A\,\ket{\Psi_A(t)}\bra{\Psi_A(t)}
\,,
\end{eqnarray}
and can be incorporated into Eq.~(\ref{neumann})
via an explicit time derivative $(\partial\hat\varrho/\partial t)_{\rm exp}$.

%%%%%%%%%%%%%%%%%%%%%%%%%%%%%%%%%%%%%%%%%%%%%%%%%%%%%%%%%%%%%%%%%%%%%%%%%%%%%%%
\subsection{Entropy}\label{entropy}
%%%%%%%%%%%%%%%%%%%%%%%%%%%%%%%%%%%%%%%%%%%%%%%%%%%%%%%%%%%%%%%%%%%%%%%%%%%%%%%

Without an explicit time-dependence 
$(\partial\hat\varrho/\partial t)_{\rm exp}$
Eq.~(\ref{neumann}) generates a unitary time evolution and hence,
the microscopic entropy remains constant in time
\begin{eqnarray}
S=-{\rm Tr}\left\{\hat\varrho\ln\hat\varrho\right\}
={\rm const}.
\end{eqnarray}
Note, that a constant microscopic entropy arises also in classical
mechanics where the time evolution is governed by the Liouville
equation. By virtue of Liouville's theorem the total time derivative
of the phase space 
density $\varrho$ vanishes and thus the classical microscopic entropy 
$\int d\Gamma\varrho\ln\varrho$ remains constant as well.  
But introducing the Boltzmann equation via averaging over
multi-particle correlations it is possible to define an effective
entropy which increases in general ($H$-theorem). 
 
An analogous procedure can be performed in quantum theory:
In practice, a complete knowledge about a given quantum
system can never be achieved.
Formally, this restriction defines a so-called observation 
level ${\cal G}=\{\hat X\}$ as a set of possibly
relevant observables $\hat X$ (see \cite{fick}), where an averaging
over all unknown and possibly irrelevant observables is understood.
With respect to a given observation level $\cal G$ one can introduce
an effective statistical operator $\hat{\varrho}_{\{\cal G\}}$ 
such that it yields the correct expectation values 
$\langle\hat X\rangle={\rm Tr}\{\hat{\varrho}_{\{\cal G\}}\hat X\}
={\rm Tr}\{\hat\varrho\hat X\}$
for all operators $\hat X\in\cal G$.
The effective statistical operator $\hat{\varrho}_{\{\cal G\}}$  
averages over all irrelevant observables $\hat X\not\in\cal G$
in order to maximize the effective entropy which is defined as
$S_{\{\cal G\}}=-{\rm Tr}
\{\hat{\varrho}_{\{\cal G\}}\ln\hat{\varrho}_{\{\cal G\}}\}$. 
This effective entropy $S_{\{\cal G\}}$ referring to a given 
observation level in general increases with time.

Introducing the internal energy $E=\langle:\hat H_0:\rangle$ 
as observation level the corresponding effective entropy
$S_E$ may increase under the influence of the dynamical external 
conditions reflecting the fact that particles have been created.
The physical meaning of $S_E$ respectively its change $\Delta S_E$
may become evident, if one assumes some energy-conserving 
relaxation process, e.g.~mediated via a measurement after the
dynamics has taken place, which thermalizes the system again 
at some higher, in principle measurable temperature $T_E=T+\Delta T$.
For a photon gas we find in the limit of high temperatures
respectively of large volumes $\mho$ the following relations between
the energy $E$, the effective entropy $S_E$ 
and the effective temperature $T_E$:
$E=\mho\,T_E^4\,\pi^2/15$ and $S_E=\mho\,T_E^3\,4\pi^4/45$. 
For small disturbances the relative increase of the effective
temperature, the energy and the effective entropy behaves as
$\Delta E/(4E)\approx\Delta T_E/T_E\approx\Delta S_E/(3S_E)$.

%%%%%%%%%%%%%%%%%%%%%%%%%%%%%%%%%%%%%%%%%%%%%%%%%%%%%%%%%%%%%%%%%%%%%%%%%%%%%%%
\subsection{Time evolution}\label{time}
%%%%%%%%%%%%%%%%%%%%%%%%%%%%%%%%%%%%%%%%%%%%%%%%%%%%%%%%%%%%%%%%%%%%%%%%%%%%%%%

Within our approach the influence of the dynamics of the external 
conditions is represented by the perturbation Hamiltonian 
$\hat H_1(t)$ governing the time-evolution of the statistical operator
in Eq.~(\ref{neumann}).
By means of the time-ordering operator ${\cal T}$ this equation can be
integrated formally  
\begin{eqnarray}
\label{time-evol}
\hat\varrho(t\uparrow\infty)
&=&
\hat U\;\hat\varrho(t\downarrow-\infty)\;\hat U^\dagger 
\nonumber\\
&=&
{\cal T}\left[\exp\left(-i\int dt'\,\hat H_1(t')\right)\right]\,
\hat\varrho(t\downarrow-\infty)
\nonumber\\
&&\times
{\cal T}^\dagger \left[\exp\left(i\int dt''\,\hat H_1(t'')\right)\right]
\,.
\end{eqnarray}
The chronological operator ${\cal T}$ acting on two bosonic and 
self-adjoint operators $\hat X(t)$ and $\hat Y(t')$ is defined by  
\begin{eqnarray}
{\cal T}\left[\hat X(t)\hat Y(t')\right]
&=&
\hat X(t)\hat Y(t')\Theta(t-t')
\nonumber\\
&&+
\hat Y(t')\hat X(t)\Theta(t'-t)
\,,
\end{eqnarray}
and so on for more operators.
Due to the Hermitian conjugation of the unitary time evolution
operator $\hat U$ in Eq.~(\ref{time-evol}), for which the position of
all operators changes, it is convenient to introduce the
anti-chronological operator  
${\cal T}^\dagger $ (cf.~\cite{schweber}) as well
\begin{eqnarray}
{\cal T}^\dagger \left[\hat X(t)\hat Y(t')\right]
&=&
\left({\cal T}\left[\hat X(t)\hat Y(t')\right]\right)^\dagger 
\nonumber\\
&=&
\hat Y(t')\hat X(t)\Theta(t-t')
\nonumber\\
&&+
\hat X(t)\hat Y(t')\Theta(t'-t)
\, .
\end{eqnarray}
Combining both equations one obtains for two bosonic and  
self-adjoint operators 
\begin{eqnarray}
\label{zeitweg}
\left\{\hat X(t),\hat Y(t')\right\}
&=&
{\cal T}\left[\hat X(t)\hat Y(t')\right]+
{\cal T}^\dagger \left[\hat X(t)\hat Y(t')\right]
\nonumber\\
&=&
\hat Y(t')\hat X(t)+
\hat X(t)\hat Y(t')
\,.
\end{eqnarray}
As we shall see below this property simplifies the calculation of the
quadratic response of the number operator.

%%%%%%%%%%%%%%%%%%%%%%%%%%%%%%%%%%%%%%%%%%%%%%%%%%%%%%%%%%%%%%%%%%%%%%%%%%%%%%%
\subsection{Canonical ensemble}\label{ensemble}
%%%%%%%%%%%%%%%%%%%%%%%%%%%%%%%%%%%%%%%%%%%%%%%%%%%%%%%%%%%%%%%%%%%%%%%%%%%%%%%

As stated in Section \ref{formalism} we assume the quantum field
to be initially at thermal equilibrium corresponding to some
non-vanishing temperature $T=1/\beta>0$. 
For reasons of simplicity we restrict our further consideration to
particles, that do not exhibit another conserved quantity than their
energy.\footnote{Otherwise, one may start with a grand canonical
ensemble.} 
This assumption is correct for photons, but not for charged
pions, for instance. We consider only bosons.
For that reason the chemical potential vanishes and the energy
$E$ is the only one observable that has a fixed expectation value.
Minimizing the microscopic entropy with this constraint generates
the canonical ensemble 
\begin{eqnarray}
\hat\varrho_0=\hat\varrho(t\downarrow-\infty)
=\frac{\exp\left(-\beta\hat H_0\right)}
{{\rm Tr}\left\{\exp\left(-\beta\hat H_0\right)\right\}}
\, .
\end{eqnarray}
There are several examples in quantum field theory where the canonical
ensemble is not capable for describing the thermal equilibrium
correctly, usually in combination with infinite volumes.
In order to treat also such cases, we assume the system to be confined 
into a finite volume where the canonical ensemble applies, calculate
the expectation values (i.e.~the trace), and consider the infinite
volume limit afterwards.
 
%%%%%%%%%%%%%%%%%%%%%%%%%%%%%%%%%%%%%%%%%%%%%%%%%%%%%%%%%%%%%%%%%%%%%%%%%%%%%%%
\subsection{Response theory}\label{response-theory}
%%%%%%%%%%%%%%%%%%%%%%%%%%%%%%%%%%%%%%%%%%%%%%%%%%%%%%%%%%%%%%%%%%%%%%%%%%%%%%%

In general, the final expectation value of an observable 
\begin{eqnarray}
\left<\hat X\,\right>
&=&
{\rm Tr}\left\{\hat X\,\hat\varrho(t\uparrow\infty)\right\}
\nonumber\\
&=&
{\rm Tr}\left\{\hat X\;\hat U\;\hat\varrho(t\downarrow-\infty)
\;\hat U^\dagger\right\}
\end{eqnarray}
cannot be calculated explicitly for non-trivial interaction terms 
$\hat H_1$ owing to the complicated structure of the corresponding 
time-evolution operator $\hat U$.
For that purpose one has usually to introduce some approximations.
One possibility is given by the perturbation expansion with respect to 
powers of the disturbance $\hat H_1$.
Assuming the perturbation Hamiltonian $\hat H_1$ to be small it is
possible to expand the above expression in powers of $\hat H_1$.
Neglecting all terms of third and higher order in $\hat H_1$ 
one obtains the quadratic response
\begin{eqnarray}
\left<\hat X\,\right>
&=&
{\rm Tr}\left\{\hat X\hat\varrho_0\right\}
+
{\rm Tr}\left\{\hat X
\left[\hat\varrho_0,i\int dt\,\hat H_1(t)\right]\right\}
\nonumber\\
&&+
{\rm Tr}\left\{
\hat X\,
\int dt\,\hat H_1(t)\,
\hat\varrho_0\,
\int dt'\,\hat H_1(t') 
\right\}
\nonumber\\
&&-
\frac{1}{2}
{\rm Tr}\left\{
\hat X\,
{\cal T}\left[\int dt\,\hat H_1(t)\int dt'\,\hat H_1(t')\right] 
\hat\varrho_0
\right\} 
\nonumber\\
&&-
\frac{1}{2}
{\rm Tr}\left\{
\hat X\,
\hat\varrho_0\,
{\cal T}^\dagger \left[\int dt\,\hat H_1(t)\int dt'\,\hat H_1(t')\right] 
\right\} 
\nonumber\\
&&+
{\cal O}\left[\hat H_1^3\right]
\, .
\end{eqnarray}
Focusing on the investigation of the particle production the relevant
observable is the number operator $\hat X=\hat N_I$.
Due to $[\hat N_I,\hat\varrho_0]=0$ the linear response vanishes 
and with the aid of Eq.~(\ref{zeitweg}) the quadratic response
simplifies to 
\begin{eqnarray}
\label{response}
\left<\hat N_I\right>
&=&
{\rm Tr}\left\{\hat N_I\hat\varrho_0\right\}
\nonumber\\
&&+
{\rm Tr}\left\{
\hat N_I\,
\left[\int dt\,\hat H_1(t),
\hat\varrho_0\right]
\int dt\,\hat H_1(t) 
\right\}
\nonumber\\
&&+
{\cal O}\left[\hat H_1^3\right]
\nonumber\\
&=&
\left<\hat N_I\right>_0
+
\Delta N_I
+
{\cal O}\left[\hat H_1^3\right]
\, .
\end{eqnarray}
The first term $\langle\hat N_I\rangle_0$ of the above expression
denotes the initial particle content in the canonical ensemble which is given 
by the Bose-Einstein distribution.
The second term $\Delta N_I$ describes the particle creation
or annihilation due to the presence of dynamical external conditions
and will be calculated in the following. 

%%%%%%%%%%%%%%%%%%%%%%%%%%%%%%%%%%%%%%%%%%%%%%%%%%%%%%%%%%%%%%%%%%%%%%%%%%%%%%%
\subsection{Particle production}\label{production}
%%%%%%%%%%%%%%%%%%%%%%%%%%%%%%%%%%%%%%%%%%%%%%%%%%%%%%%%%%%%%%%%%%%%%%%%%%%%%%%

For quasi-free quantum fields obeying linear equations of motion the
perturbation Hamiltonian can be expressed -- up to an irrelevant
constant -- as a bilinear form of the fields.
Expanding the fields into the time-independent creation/annihilation 
operators representing particles in the asymptotic regions
the disturbance $\hat H_1$ can be cast into the rather general form 
\begin{eqnarray}
\label{allgH1}
\int dt\,\hat H_1(t)
&=&
\frac{1}{2}
\left(
{\cal S}_{{J}{K}}\hat a_{J}^\dagger \hat a_{K}^\dagger 
+
{\cal S}_{{J}{K}}^*\hat a_{J}\hat a_{K}
\right)
\nonumber\\
&&+
{\cal U}_{{J}{K}}\hat a_{J}^\dagger \hat a_{K}
+
{\cal C}
\,.
\end{eqnarray}
The introduced matrices have to fulfill the conditions 
${\cal S}_{{J}{K}}={\cal S}_{{K}{J}}$ and 
${\cal U}_{{J}{K}}={\cal U}_{{K}{J}}^*$
because $\hat H_1$ is self-adjoint.
The ${\cal S}$-term could be interpreted as a generator
for a multi-mode squeezing operator and the 
${\cal U}$-term as a hopping operator.
Now the matrices ${\cal S}$ and ${\cal U}$ 
contain all information about the dynamical external conditions that
are relevant for the quadratic response. 
Inserting the general form of the disturbance in Eq.~(\ref{allgH1})
into Eq.~(\ref{response}) after evaluating the traces the quadratic
response of the number operator takes the form
\begin{eqnarray}
\label{deltaN}
\Delta N_I
&=&
\left|{\cal S}_{I{J}}\right|^2
\left(1+\left<\hat N_{J}\right>_0+\left<\hat N_I\right>_0\right)
\nonumber\\
&&+
\left|{\cal U}_{I{J}}\right|^2
\left(\left<\hat N_{J}\right>_0-\left<\hat N_I\right>_0\right)
\, .
\end{eqnarray}
Note, that there is still a summation over the index ${J}$.
One observes that merely the $\cal S$-term governs the production of 
particles and contains the vacuum contribution (first addend).
The $\cal U$-term does not increase the total number of particles
since it has the same structure as a classical master-equation
(e.g.~used for the derivation of the $H$-theorem), 
but it transforms particles from one mode into another and 
thereby also increases the energy.    
Investigating the high-temperature expansion of the Bose-Einstein
distribution entering the equation above
\begin{eqnarray}
\label{highT}
\left<\hat N_I\right>_0
=
\frac{1}{\exp(\beta\omega_I)-1}=
\frac{1}{\beta\omega_I}-\frac{1}{2}+{\cal O}[\beta]
\end{eqnarray}
one observes that the temperature-independent contributions 
(the term $-1/2$) cancel.
The same occurs in the static Casimir effect
\cite{plunien,balian,mostepanenko} and may be interpreted 
as a consequence of the scale invariance in the classical limit.
In the high-temperature limit the expectation value 
$\langle\hat N_I\rangle_0$ is linear in $T$.
But special care is required for evaluating the number of
created particles in Eq.~(\ref{deltaN}) due to the remaining
mode summation.
Since the expansion in Eq.~(\ref{highT}) has a finite
radius of convergence its insertion into Eq.~(\ref{deltaN})
may cause some problems in performing the mode summation.
Therefore the resulting number of produced particles or the
total radiated energy may possibly display another behavior
as shown in Eq.~(\ref{highT}).
As we shall see later in the Sections \ref{moving} and \ref{smallR}, 
the radiated energy may be proportional to even higher powers in $T$.       

%%%%%%%%%%%%%%%%%%%%%%%%%%%%%%%%%%%%%%%%%%%%%%%%%%%%%%%%%%%%%%%%%%%%%%%%%%%%%%%
\subsection{Correlations}\label{correlations}
%%%%%%%%%%%%%%%%%%%%%%%%%%%%%%%%%%%%%%%%%%%%%%%%%%%%%%%%%%%%%%%%%%%%%%%%%%%%%%%

The expectation value $\langle\hat N_I\rangle$ of the 
number operator after the dynamical period does in general not
display the Bose-Einstein distribution because the field is no
longer in the thermodynamic equilibrium.
But for certain dynamics this expectation value could still behave 
as a thermal distribution corresponding to some effective 
temperature $T_{\rm eff}$. 
If particles are created during the dynamics this effective
temperature will be larger than the initial temperature $T$.
Such a phenomenon may occur even for a vanishing initial temperature 
$T=0$. 
This effect can be explained within the thermo-field  
\cite{thermo-field1,thermo-field2} formalism:
Measuring only single-particle observables does not reveal the complete
information about the quantum system under consideration.
Formally, this restriction defines an observation level $\cal G$
\cite{fick} including all these single-particle observables 
(see the remarks in the previous Section).
The effective density matrix $\hat{\varrho}_{\{{\cal G}\}}$ may indeed
be equal to the statistical operator of a canonical ensemble
corresponding to some effective temperature $T_{\rm eff}$.
However, the real density matrix $\hat\varrho$ cannot be equal to this 
effective statistical operator $\hat{\varrho}_{\{{\cal G}\}}$ because the
microscopic entropy $S=-{\rm Tr}\left\{\hat\varrho\ln\hat\varrho\right\}$ is
conserved during a unitary time-evolution while the effective entropy   
$S_{\{\cal G\}}=-{\rm Tr}\{\hat{\varrho}_{\{\cal G\}}
\ln\hat{\varrho}_{\{\cal G\}}\}$ 
has been increased for $T<T_{\rm eff}$.
Within the investigation of only single-particle observables one can never
distinguish between the two statistical operators $\hat\varrho$ and
$\hat{\varrho}_{\{{\cal G}\}}$ and therefore one can never find out whether
the measured temperature represents a real one ($T$) or an effective one 
($T_{\rm eff}$). 
For this purpose it is necessary to consider many-particle observables.
One suitable candidate is given by the two-particle correlation defined as
\begin{eqnarray}
{\cal C}_{{J}{K}}=\left<\hat N_{J}\hat N_{K}\right>
-\left<\hat N_{J}\right>
\left<\hat N_{K}\right>
\quad{\rm for}\quad{J}\neq{K}
\,.
\end{eqnarray}
This quantity is particularly appropriate since it
vanishes in the thermodynamic equilibrium
\begin{eqnarray}
\left<\hat N_{J}\hat N_{K}\right>_0=
\left<\hat N_{J}\right>_0
\left<\hat N_{K}\right>_0
\quad{\rm for}\quad{J}\neq{K}
\,.
\end{eqnarray}
The quadratic response of the correlation function 
can be evaluated as follows
\begin{eqnarray}
\label{correl}
{\cal C}_{{J}{K}}
&=&
\left<\int dt\,\hat H_1(t)
\left[\hat N_{J}\hat N_{K},\int dt\,\hat H_1(t)\right]\right>_0
\nonumber\\
&&-\left<\hat N_{J}\right>_0\Delta N_{K}
-\left<\hat N_{K}\right>_0\Delta N_{J}
+{\cal O}\left[\hat H_1^3\right]
\,.
\end{eqnarray}
As a simple example we may consider the vacuum case with $T=0$
\begin{eqnarray}
{\cal C}_{{J}{K}}
&=&
\bra{0}\int dt\,\hat H_1(t)\hat N_{J}\hat N_{K}\int dt\,
\hat H_1(t)\ket{0} 
\nonumber\\
&&+
{\cal O}\left[\hat H_1^3\right]
\,,
\end{eqnarray}
where the correlation is positive 
(at least in lowest order in $\hat H_1$).

But at finite temperatures the correlation may also
assume negative values:
E.g., in the case of a completely diagonal perturbation Hamiltonian we 
arrive at 
\begin{eqnarray}
{\cal C}_{{J}{K}}=
-\left<\hat N_{J}\right>_0\Delta N_{K}
-\left<\hat N_{K}\right>_0\Delta N_{J}
+{\cal O}\left[\hat H_1^3\right]
\,.
\end{eqnarray}
%

%%%%%%%%%%%%%%%%%%%%%%%%%%%%%%%%%%%%%%%%%%%%%%%%%%%%%%%%%%%%%%%%%%%%%%%%%%%%%%%
\subsection{Bogoliubov transformation}\label{transformation}
%%%%%%%%%%%%%%%%%%%%%%%%%%%%%%%%%%%%%%%%%%%%%%%%%%%%%%%%%%%%%%%%%%%%%%%%%%%%%%%

It might be illuminating to consider the relation of the previous
investigations to the formalism based on the Bogoliubov transformation.
The initial and final creation and annihilation operators are
connected through the Bogoliubov coefficients via
\begin{eqnarray}
\hat U^\dagger\hat a_{J}\hat U=
\alpha_{{J}{K}}\hat a_{K}+\beta_{{J}{K}}\hat a_{K}^\dagger
\,.
\end{eqnarray}
Switching from the interaction picture to the Heisenberg
representation the expectation value of the number operator
\begin{eqnarray}
\left<\hat N_{J}\right>={\rm Tr}\left\{\hat\varrho\hat N_{J}\right\}
&=&
{\rm Tr}\left\{\hat\varrho_0\hat U^\dagger\hat N_{J}\hat U\right\}
\nonumber\\
&=&
\left<\hat U^\dagger\hat N_{J}\hat U\right>_0
\end{eqnarray}
can be expressed in terms of the Bogoliubov coefficients
\begin{eqnarray}
\left<\hat N_{J}\right>=
\left|\beta_{{J}{K}}\right|^2+
\left<\hat N_{K}\right>_0
\left(\left|\alpha_{{J}{K}}\right|^2+\left|\beta_{{J}{K}}\right|^2\right)
\,.
\end{eqnarray}
If we assume a completely diagonal Hamiltonian 
$\hat H=\hat H_0+\hat H_1$ the associated 
time-evolution operator $\hat U$ factorizes
\begin{eqnarray}
\hat U=\prod\limits_I\hat U_I
\,.
\end{eqnarray}
Accordingly the Bogoliubov coefficients simplify to
$\alpha_{{J}{K}}=\alpha_{{J}}\delta_{{J}{K}}$ and 
$\beta_{{J}{K}}=\beta_{{J}}\delta_{{J}{K}}$, respectively.
Utilizing the unitary relation, which assumes for the
diagonal coefficients the simple form   
$|\alpha_I|^2=1+|\beta_I|^2$, we arrive at
\begin{eqnarray}
\left<\hat N_I\,\right>=\left<\hat N_I\,\right>_0+
|\beta_I|^2\left(1+2\left<\hat N_I\,\right>_0\right)
\, .
\end{eqnarray}
Hence -- for a diagonal Hamiltonian -- the number of created particles
at finite temperature is simply given by the corresponding vacuum
expression times a thermal factor
\begin{eqnarray}
\Delta N_\omega^{T}=
\Delta N_\omega^{T=0}
\left(1+\frac{2}{\exp(\beta\omega)-1}\right)
\, .
\end{eqnarray}
It should be mentioned here that this result is not restricted to 
a particular order perturbation theory -- it holds for the general
case of a diagonal Hamiltonian.

%%%%%%%%%%%%%%%%%%%%%%%%%%%%%%%%%%%%%%%%%%%%%%%%%%%%%%%%%%%%%%%%%%%%%%%%%%%%%%%
%%%%%%%%%%%%%%%%%%%%%%%%%%%%%%%%%%%%%%%%%%%%%%%%%%%%%%%%%%%%%%%%%%%%%%%%%%%%%%%
\section{Trembling cavities}\label{trembling}
%%%%%%%%%%%%%%%%%%%%%%%%%%%%%%%%%%%%%%%%%%%%%%%%%%%%%%%%%%%%%%%%%%%%%%%%%%%%%%%
%%%%%%%%%%%%%%%%%%%%%%%%%%%%%%%%%%%%%%%%%%%%%%%%%%%%%%%%%%%%%%%%%%%%%%%%%%%%%%%

Now we are in the position to apply the formalism presented in the
previous Section to a special system of a quantum field under the 
influence of dynamical external conditions.
We consider a massless and neutral scalar field $\Phi$ confined in an
arbitrary and weakly time-dependent domain $G(t)$ (a trembling cavity)
with Dirichlet boundary conditions $\Phi=0$ at $\partial G(t)$, see
also Refs.~\cite{trembling} and \cite{diplom}. 
The action generating the equation of motion is given by 
\begin{eqnarray}
\label{tremaction}
{\cal A}=\frac12\int dt \int\limits_G d^3r
(\partial_\mu\Phi)(\partial^\mu\Phi)=\int dt\,L\,.
\end{eqnarray}
The Hamiltonian governing the dynamics of this system can be obtained
in the following way:
For every fixed time $t$ we construct a complete and orthonormal set
of {\em eigen\,}functions of the Laplacian inside the time-dependent
domain $G(t)$ satisfying the required Dirichlet boundary conditions.
Owing to the time-dependence of the Dirichlet boundary conditions at
$\partial G(t)$ this operator becomes time-dependent as well.
Hence also the proper {{\em eigen\,}}functions are time-dependent
$f_I=f_I(t)$. 
We assume a finite domain $G(t)$ resulting in a purely discrete
spectrum. 
The insertion of the expansion of the field
\begin{eqnarray}
\label{expandphi}
\hat\Phi(t,\f{r})=\sum\limits_I \hat Q_I(t) f_I(t,\f{r})
\end{eqnarray}
into the Lagrangian in Eq.~(\ref{tremaction}) reveals 
\begin{eqnarray}
L
&=&
\frac{1}{2}\left(\dot Q_I^2-\omega_I^2(t)\,Q_I^2\right)+
Q_I\,{\cal M}_{IJ}(t)\,\dot Q_J
\nonumber\\
&&+
\frac{1}{2}Q_I\,{\cal M}_{IJ}(t)\,{\cal M}_{JK}(t)\,Q_K
\,.
\end{eqnarray}
Since the time-derivative of the field $\Phi$ may also include the
explicit time-derivative of the {{\em eigen\,}}functions, we obtain an 
additional anti-symmetric inter-mode coupling matrix
\begin{eqnarray}
{\cal M}_{IJ}(t)=\int\limits_{G(t)} d^3r\,f_I\,\dot f_J\,.
\end{eqnarray}
Furthermore, the {{\em eigen\,}}values of the Laplace operator 
$\omega_I^2(t)$ are time-dependent in general.
By means of the Legendre transformation we obtain the Hamiltonian
\begin{eqnarray}
H=
\frac{1}{2}\left(P_I^2-\omega_I^2(t)\,Q_I^2\right)+
Q_I\,{\cal M}_{IJ}(t)\,P_J
\,,
\end{eqnarray}
where $P_{J}$ and $Q_{K}$ are the canonical conjugated variables.
In the following the undisturbed {{\em eigen\,}}values (frequencies)
are denoted 
by $\omega_{I}^2(|t|\uparrow\infty)=\omega_{I}^2$ and their variation
by $\Delta\omega_{I}^2(t)=\omega_{I}^2(t)-\omega_{I}^2$.
After the canonical quantization
and the separation of the undisturbed part 
\begin{eqnarray}
\hat H_0
=\frac{1}{2}\left(\hat P_{J}^2+\omega_{J}^2\hat Q_{J}^2 \right)
\end{eqnarray}
the perturbation Hamiltonian may be cast into the form
\begin{eqnarray}
\label{cavH1}
\hat H_1(t)
&=&
\Delta\hat E(t)+\hat W(t)
\nonumber\\
&=&
\frac{1}{2}\hat Q_{J}^2(t)\Delta\omega_{J}^2(t)
+\hat Q_{J}(t){\cal M}_{{J}{K}}(t)\hat P_{K}(t)
\,.
\end{eqnarray}
The term including 
$\Delta\omega_{J}^2(t)$, i.e.~$\Delta\hat E(t)$,
arises from the change of the shape of the domain $G(t)$.
This term $\Delta\hat E(t)$ is called squeezing contribution. 
The inter-mode coupling ${\cal M}_{{J}{K}}(t)$ contained in $\hat W(t)$ 
-- the velocity contribution -- arises from the motion of the 
boundaries.
Of course, the undisturbed Hamiltonian can be diagonalized through a
particle definition 
$\hat H_0=\omega_{J}(\hat N_{J}+1/2)$
employing the usual creation/annihilation operators
\begin{eqnarray}
\hat a_{J}=\sqrt{\frac{1}{2\omega_{J}}}
\left(\omega_{J}\hat Q_{J}+i\hat P_{J}\right)
\, .
\end{eqnarray}
With the aid of the equations above it is now possible to 
calculate the expectation value of the number operator
employing the results of the previous Section.
The evaluation of the trace $\rm Tr\{\dots\}$ is most
suitably performed in the basis of the 
$\hat H_0$-{{\em eigen\,}}kets.
One obtains a non-vanishing trace only for those terms that 
contain the same number of creation and 
annihilation operators for every mode.
Inserting the particular interaction Hamiltonian
$\hat H_1(t)=\Delta\hat E(t)+\hat W(t)$ into Eq.~(\ref{response})
generates at a first glance four terms.
However, owing to the antisymmetry of the matrix governing the 
inter-mode coupling ${\cal M}_{{J}{J}}=0$ the mixing terms vanish
\begin{eqnarray}
{\rm Tr}\left\{
\hat N_I\,
\left[\Delta\hat E(t),
\hat\varrho_0\right]
\hat W(t') 
\right\}
&=&0\,,
\nonumber\\
{\rm Tr}\left\{
\hat N_I\,
\left[\hat W(t'),
\hat\varrho_0\right]
\Delta\hat E(t) 
\right\}
&=&0
\,.
\end{eqnarray}
Consequently the squeezing and the velocity contribution
decouple on the level of the quadratic response
\begin{eqnarray}
\Delta N_I
&=&
{\rm Tr}\left\{
\hat N_I\,
\left[\int dt\,\Delta\hat E(t),
\hat\varrho_0\right]
\int dt\,\Delta\hat E(t) 
\right\}
\nonumber\\
&&+
{\rm Tr}\left\{
\hat N_I\,
\left[\int dt\,\hat W(t),
\hat\varrho_0\right]
\int dt\,\hat W(t) 
\right\}
\nonumber\\
&=&
\Delta N_I^S+\Delta N_I^V
\, .
\end{eqnarray}
Expressing $\hat H_1(t)=\Delta\hat E(t)+\hat W(t)$ with the aid
of the matrices $\cal S$ and $\cal U$ as done in Eq.~(\ref{allgH1}) 
one observes that the squeezing effect manifests in the diagonal 
elements of the matrices $\cal S$ and $\cal U$ while the velocity 
effect generates off-diagonal elements only.

%%%%%%%%%%%%%%%%%%%%%%%%%%%%%%%%%%%%%%%%%%%%%%%%%%%%%%%%%%%%%%%%%%%%%%%%%%%%%%%
\subsection{Squeezing}\label{squeezing}
%%%%%%%%%%%%%%%%%%%%%%%%%%%%%%%%%%%%%%%%%%%%%%%%%%%%%%%%%%%%%%%%%%%%%%%%%%%%%%%

The diagonal form of the squeezing $\Delta\hat E(t)$-part of the
perturbation Hamiltonian $\hat H_1(t)$ indicates the
highly resonant character of this contribution.
This fact simplifies the calculation of $\Delta N_I$ because 
only one mode -- the mode $I$ -- survives in the trace. 
With the abbreviation
\begin{eqnarray}
\label{fourier}
{\cal Q}_I^S
&=&
\frac{1}{2\omega_I}
\int dt\,\Delta\omega_I^2(t)
\exp\left(2i\omega_I t\right)
\nonumber\\
&=&
\frac{1}{2\omega_I}
\widetilde{\Delta\omega_I^2\,}(2\omega_I)
\end{eqnarray}
the squeezing part ${\cal S}^S$ of the matrix 
${\cal S}={\cal S}^S+{\cal S}^V$ can be cast into the form
\begin{eqnarray}
{\cal S}_{{J}{K}}^S={\cal Q}_I^S
\,\delta_{I{J}}\,\delta_{I{K}}
\, .
\end{eqnarray}
Of course, the squeezing part ${\cal U}^S$ of the matrix 
${\cal U}={\cal U}^S+{\cal U}^V$ is also
strictly diagonal and hence does not contribute to $\Delta N_I$
(see Eq.~(\ref{response})) with the result that only the ${\cal S}^S$ 
term is responsible for particle production.
Therefore the squeezing contribution to the number of particles reads
\begin{eqnarray}
\Delta N_I^S
&=&
\left|{\cal Q}_I^S\right|^2
\left(
1+2\left<\hat N_I\right>_0
\right)
\nonumber\\
&=&
\left|{\cal Q}_I^S\right|^2
\left(
1+\frac{2}{\exp(\beta\omega_I)-1}
\right)
\, .
\end{eqnarray}
In accordance to the results of Sec.~\ref{transformation} the particle
production  
rate at temperature $T$ equals the rate at zero temperature times the 
thermal distribution factor. 

%%%%%%%%%%%%%%%%%%%%%%%%%%%%%%%%%%%%%%%%%%%%%%%%%%%%%%%%%%%%%%%%%%%%%%%%%%%%%%%
\subsection{Velocity}\label{velocity}
%%%%%%%%%%%%%%%%%%%%%%%%%%%%%%%%%%%%%%%%%%%%%%%%%%%%%%%%%%%%%%%%%%%%%%%%%%%%%%%

Due to the more complicated structure of the velocity term,
i.e.~the off-diagonal elements, the calculation of the number
of created particles involves an additional summation.
Hence the velocity contribution may not be cast into such
a simple form as the squeezing term.
The $\hat W$-part of the perturbation Hamiltonian can be
expanded with the aid of the matrices
\begin{eqnarray}
{\cal S}_{{J}{K}}^V
&=&
\frac{i}{2}\int dt\,{\cal M}_{{J}{K}}(t)\,
\exp\left(i[\omega_{J}+\omega_{K}]t\right)
\nonumber\\
&&\times
\left(\sqrt{\frac{\omega_{J}}{\omega_{K}}}-
\sqrt{\frac{\omega_{K}}{\omega_{J}}}\right)
\,,
\end{eqnarray}
and
\begin{eqnarray}
{\cal U}_{{J}{K}}^V
&=&
\frac{i}{2}\int dt\,{\cal M}_{{J}{K}}(t)\,
\exp\left(i[\omega_{J}-\omega_{K}]t\right)
\nonumber\\
&&\times
\left(\sqrt{\frac{\omega_{J}}{\omega_{K}}}+
\sqrt{\frac{\omega_{K}}{\omega_{J}}}\right)
\, .
\end{eqnarray}
In this case both terms, the ${\cal S}^V$ and the ${\cal U}^V$ matrices
contribute.
The ${\cal U}^V$ term may even decrease the number of particles in a 
given mode, see Eq.~(\ref{response}). For all that the total energy
increases.

%%%%%%%%%%%%%%%%%%%%%%%%%%%%%%%%%%%%%%%%%%%%%%%%%%%%%%%%%%%%%%%%%%%%%%%%%%%%%%%
\subsection{Moving mirror}\label{moving}
%%%%%%%%%%%%%%%%%%%%%%%%%%%%%%%%%%%%%%%%%%%%%%%%%%%%%%%%%%%%%%%%%%%%%%%%%%%%%%%

In order to illustrate the velocity effect we consider the 
most simple example of a single mirror in 1+1 dimensions.
This scenario has already been investigated by several authors, 
e.g.~\cite{fulling+davies,ford+vilenkin}, at zero temperature. 
In this case the domain $G$ takes the form $G(t)=(\eta(t),\infty)$
where $\eta(t)$ denotes the time-dependent position of the mirror
with $\eta(t\downarrow-\infty)=\eta(t\uparrow+\infty)=0$.
The index $I$ can be identified with the wave number 
$I\rightarrow k=\omega_{I}$ which assumes all positive real numbers. 
As the shape of the domain $G(t)$ does not change only
the velocity-effect contributes. 
The inter-mode coupling matrix is given by 
(cf.~\cite{trembling,diplom}) 
\begin{eqnarray}
{\cal M}_{IJ}(t)
\rightarrow
{\cal M}_{kk'}(t)
=
\dot\eta(t)\frac{2}{\pi}
{\cal P}\left[\frac{kk'}{k^2-k'^2}\right]
\,,
\end{eqnarray}
where ${\cal P}$ denotes the principal value.
The Fourier transform of a time-dependent function 
is denoted by a tilde: $\widetilde{\eta}={\cal F}\eta$
(see Eq.~(\ref{fourier})).
Using this notation the matrices for the evaluation of
the particle number read
\begin{eqnarray}
{\cal S}_{kk'}
=
\frac{1}{\pi}
\left(\sqrt{\frac{k}{k'}}-\sqrt{\frac{k'}{k}}\right)
\frac{kk'}{k-k'}\,\widetilde{\eta}(k+k')
\end{eqnarray}
and
\begin{eqnarray}
{\cal U}_{kk'}
=
\frac{1}{\pi}
\left(\sqrt{\frac{k}{k'}}+\sqrt{\frac{k'}{k}}\right)
\frac{kk'}{k+k'}\,\widetilde{\eta}(k-k')
\, .
\end{eqnarray}
Inserting these expressions into Eq.~(\ref{deltaN}) yields
the following results for the expectation values of the 
number operator 
\begin{eqnarray}
\label{radpart}
\Delta N_k
&=&
\int\limits_0^\infty dk'\,
\frac{kk'}{\pi^2}\left|\widetilde{\eta}(k+k')\right|^2
\left(1+\left<\hat N_{k'}\right>_0+\left<\hat N_{k}\right>_0\right)
\nonumber\\
&&+
\int\limits_0^\infty dk'\,
\frac{kk'}{\pi^2}\left|\widetilde{\eta}(k-k')\right|^2
\left(\left<\hat N_{k'}\right>_0-\left<\hat N_k\right>_0\right)
\,.
\nonumber\\
\end{eqnarray}
Already for the most simple example the velocity contribution cannot
be cast into a form being as simple as the squeezing term.
But the formula above allows us to calculate the number of created
particles within the quadratic response for arbitrary dynamics 
$\eta(t)$ and temperatures $T$.
Deriving the total radiated energy from
Eq.~(\ref{radpart})  
\begin{eqnarray}
\label{raden}
E=\int\limits_0^\infty dk\,k\,\Delta N_k
=\int dt\int dt'\,\eta(t)\,\eta(t')\,{\cal R}(t-t')
\,,
\end{eqnarray}
where ${\cal R}$ denotes the quadratic response function,
we obtain a more elucidative formula.
For that purpose we have to perform integrations involving 
Bose-Einstein distribution functions entering in 
$\langle\hat N_k\rangle_0$.
If we insert the usual expansion for those functions
\begin{eqnarray}
\left<\hat N_k\right>_0
=
\frac{1}{\exp(\beta k)-1}
=
\sum\limits_{n=1}^{\infty}\,e^{-n\beta k}
\end{eqnarray}
the wave number integration $I^m_\beta$ leads to Hurwitz
zeta-functions 
\begin{eqnarray}
I^m_\beta
&=&
\sum\limits_{n=1}^{\infty}\int\limits_0^\infty dk\,k^m
\exp\left(ik(t-t')-n\beta k\right)
\nonumber\\
&=&
\sum\limits_{n=1}^{\infty}\frac{m!}{\left(n\beta-i[t-t']\right)^{m+1}}
\nonumber\\
&=&
\frac{m!}{\beta^{m+1}}\,\zeta\left(m+1,1-i\frac{t-t'}{\beta}\right)
\, .
\end{eqnarray}
In terms of these functions the response function 
${\cal R}={\cal R}(\Delta t)={\cal R}(t-t')$ yields
after the $k$- and $k'$-integrations 
\begin{eqnarray}
{\cal R}
&=&
\frac{2}{\pi^2}\left(
+\pi\delta^{(4)}(\Delta t)+
{\cal P}\left[\frac{24i}{\Delta t^5}\right]
\right)
\frac{1}{24}
\nonumber\\
&&+
\frac{1}{\pi^2}\left(-\pi\delta^{(2)}(\Delta t)
-{\cal P}\left[\frac{2i}{\Delta t^3}\right]\right)
\frac{\zeta(2,1-i\Delta t/\beta)}{\beta^2}
\nonumber\\
&&+
\frac{2}{\pi^2}\left(-i\pi\delta^{(1)}(\Delta t)
-{\cal P}\left[\frac{1}{\Delta t^2}\right]\right)
\frac{\zeta(3,1-i\Delta t/\beta)}{\beta^3}
\nonumber\\
&&+
\frac{1}{\pi^2}\left(-\pi\delta^{(2)}(\Delta t)
+{\cal P}\left[\frac{2i}{\Delta t^3}\right]\right)
\frac{\zeta(2,1-i\Delta t/\beta)}{\beta^2}
\nonumber\\
&&+
\frac{2}{\pi^2}\left(-i\pi\delta^{(1)}(\Delta t)
+{\cal P}\left[\frac{1}{\Delta t^2}\right]\right)
\frac{\zeta(3,1-i\Delta t/\beta)}{\beta^3}
\,.
\nonumber\\
\end{eqnarray}
The five terms above correspond directly to the five terms in
Eq.~(\ref{radpart}).
As one can easily check in Eq.~(\ref{raden}), only the symmetric part
${\cal R}_{\rm sym}(\Delta t)=({\cal R}(\Delta t)+{\cal R}(-\Delta t))/2$ 
of the response function ${\cal R}(\Delta t)$ contributes to the
radiated energy.
Symmetrizing the response function a lot of cancellations occur and
all divergent terms of the structure $1/\Delta t^n$ disappear. 
The resulting expression reads
\begin{eqnarray}
{\cal R}_{\rm sym}(\Delta t)
=
\frac{1}{12\pi}\delta^{(4)}(\Delta t)
-
\frac{2}{\pi}\zeta(2)T^2\delta^{(2)}(\Delta t)
\,,
\end{eqnarray}
with the Riemann zeta-function $\zeta(n)$ that is related to the
Hurwitz zeta-function $\zeta(n,m)$ via $\zeta(n,1)=\zeta(n)$.
Rewriting this expression into the total radiated energy yields
\begin{eqnarray}
E=\frac{1}{12\pi}\int dt\,\ddot\eta^2(t)+
\frac{\pi}{3}T^2\int dt\,\dot\eta^2(t)
\, .
\end{eqnarray}
The first term describes the vacuum contribution and was originally 
obtained by Fulling and Davies \cite{fulling+davies} using the conformal
invariance of the scalar field in 1+1 dimensions and has been later
calculated by Ford and Vilenkin \cite{ford+vilenkin} via a more flexible
method of perturbations of Green functions.
In both approaches the radiated energy was deduced by means of the
point-splitting renormalization technique.  
The relevant contributions of the Green functions used in 
Refs.~\cite{fulling+davies,ford+vilenkin} correspond to the
vacuum part of the response function ${\cal R}(\Delta t)$ in our
derivation.  

The second term is a pure temperature effect and generalizes the 
vacuum results in Refs.~\cite{fulling+davies,ford+vilenkin} 
to the density matrix corresponding to the canonical ensemble.   
The relation between the finite temperature correction 
to the radiated energy and the vacuum contribution
is of the order ${\cal O}[T^2\tau^2]$ where $\tau$ denotes 
a characteristic time scale of the dynamics of the mirror. 

%%%%%%%%%%%%%%%%%%%%%%%%%%%%%%%%%%%%%%%%%%%%%%%%%%%%%%%%%%%%%%%%%%%%%%%%%%%%%%%
%%%%%%%%%%%%%%%%%%%%%%%%%%%%%%%%%%%%%%%%%%%%%%%%%%%%%%%%%%%%%%%%%%%%%%%%%%%%%%%
\section{Resonantly vibrating cavity}\label{resonantly}
%%%%%%%%%%%%%%%%%%%%%%%%%%%%%%%%%%%%%%%%%%%%%%%%%%%%%%%%%%%%%%%%%%%%%%%%%%%%%%%
%%%%%%%%%%%%%%%%%%%%%%%%%%%%%%%%%%%%%%%%%%%%%%%%%%%%%%%%%%%%%%%%%%%%%%%%%%%%%%%

Let us now investigate the finite-temperature effects on the dynamical
Casimir-effect in a resonantly vibrating cavity.
In order to allow for an experimental verification
the number of motion-induced particles should be as large as possible.
One way to achieve this goal is to exploit of the phenomenon
of parametric resonance.
It occurs in the case of periodically time-dependent perturbations 
characterized by some frequency $\omega$.
Within the quadratic response the number $\Delta N_I$ of created 
particles is proportional to the Fourier transform of the perturbation 
function (see Eq.~(\ref{fourier})).
Assuming a harmonically oscillating disturbance the Fourier transform
possesses a pronounced maximum at the resonance frequency. 
As a result particles with a mode frequency corresponding to $\omega$ will
be produced predominantly.
Obtaining large numbers may indicate that one has left the region,
where second-order perturbation theory does apply.

%%%%%%%%%%%%%%%%%%%%%%%%%%%%%%%%%%%%%%%%%%%%%%%%%%%%%%%%%%%%%%%%%%%%%%%%%%%%%%%
\subsection{Rotating wave approximation}\label{rotating} 
%%%%%%%%%%%%%%%%%%%%%%%%%%%%%%%%%%%%%%%%%%%%%%%%%%%%%%%%%%%%%%%%%%%%%%%%%%%%%%%

In the case of oscillating disturbances, however, it is possible to evaluate 
the time evolution operator $\hat U$ in all orders of $\hat H_1$ 
analytically employing yet another approximation, 
the so-called rotating wave approximation 
(RWA, see e.g.~\cite{law}).
The main consequence of the RWA consists in the fact that it allows 
for the derivation of a time-independent effective Hamiltonian
$\hat H_1^{\rm eff}$ after performing the integration over time
\begin{eqnarray}
\int dt\,\hat H_1\approx{\sf T}\,\hat H_1^{\rm eff}
\, .
\end{eqnarray}
Let us assume that the explicit time-dependence of the perturbation 
Hamiltonian $\hat H_1(t)$ possesses an oscillatory behavior like
$\varepsilon\sin(2\omega t)$ during a  sufficiently long time {\sf T},
such that the conditions $\omega{\sf T}\gg 1$, $\varepsilon\ll 1$ and 
$\varepsilon\,\omega{\sf T}={\cal O}[1]$ hold.
Expanding the time evolution operator $\hat U$ into powers of
$\varepsilon$ and $\omega{\sf T}$ the RWA neglects all terms of order
${\cal O}[\varepsilon^n(\omega{\sf T})^m]$ if $n>m$ holds.
Since time integrations over oscillating functions result in smaller
powers of ${\sf T}$ than the same integrations over time-independent
quantities, within the RWA all terms including oscillations were omitted.
Accordingly, only those terms, where the oscillations due to the
time-dependence of the operators 
(in the interaction picture governed by $\hat H_0$)
and the explicit time-dependent disturbances cancel -- i.e.~which are 
in resonance ($n=m$) -- contribute in the RWA.
This approximation enables us to neglect the time-ordering $\cal T$ as
well. 
The difference between the time-ordered and the original expression
always contains commutators like $[\hat H_1(t),\hat H_1(t')]$.
These quantities are always oscillating and therefore can be neglected 
within the RWA.

%%%%%%%%%%%%%%%%%%%%%%%%%%%%%%%%%%%%%%%%%%%%%%%%%%%%%%%%%%%%%%%%%%%%%%%%%%%%%%%
\subsection{Fundamental resonance}\label{fundamental}
%%%%%%%%%%%%%%%%%%%%%%%%%%%%%%%%%%%%%%%%%%%%%%%%%%%%%%%%%%%%%%%%%%%%%%%%%%%%%%%

For a harmonically vibrating cavity the effective Hamiltonian 
$\hat H_1^{\rm eff}$
can easily be calculated from the interaction operator in
Eq.~(\ref{cavH1}). 
Assuming harmonic time-dependences $\sim\varepsilon\sin(2\omega t)$
or $\sim\varepsilon\cos(2\omega t)$ for both, the squeezing
($\Delta\omega_{I}^2(t)$) and the velocity terms (${\cal M}_{{J}{K}}(t)$),
only those terms will survive, which match the resonance conditions.
These conditions are fulfilled if the oscillations of the
operators, i.e.~$\hat Q_{J}(t)$ and $\hat P_{K}(t)$, compensate the
oscillations of the disturbances, i.e.~$\Delta\omega_{I}^2(t)$ (squeezing) 
and ${\cal M}_{{J}{K}}(t)$ (velocity).
For the squeezing term the resonance condition reads 
$\omega_I=\omega$ and for the velocity term 
$|\omega_{J}\pm\omega_{K}|=2\omega$, respectively.
Accordingly, the squeezing effect creates always particles
with the frequency $\omega$ provided this cavity mode does exist.
We restrict our further consideration to the situation, where the
oscillation frequency $\omega$ corresponds to the lowest cavity mode,
i.e.~to the fundamental resonance
\begin{eqnarray}
\omega
={\rm min}\left\{\omega_I\right\}
=\omega_1
\,.
\end{eqnarray}
The fundamental resonance frequency $\omega_1$ is determined by the
characteristic size $\Lambda$ of the cavity, e.g.~
$\omega_1=\sqrt{3}\pi/\Lambda$ for a cubic cavity.
For the lowest mode $I=1$ the resonance condition for squeezing
$\omega_I=\omega_1=\omega$ is satisfied automatically.
Although the condition for the $\cal S$-term of the velocity effect
$|\omega_{J}+\omega_{K}|=2\omega$ could be satisfied for
${J}={K}=1$ (we assume a non-degenerate ground state $I=1$),
this term does not contribute since ${\cal M}_{{J}{J}}=0={\cal S}_{{J}{J}}$.
Whether the resonance condition for the $\cal U$-term of the velocity effect
$|\omega_{J}-\omega_{K}|=2\omega$
can be satisfied or not depends on the spectrum of the particular
cavity under consideration.
For a one-dimensional cavity the {{\em eigen\,}}frequencies $\omega_I$
are proportional to integers and thus it can be satisfied leading to
an additional velocity contribution.
For most cases of higher-dimensional cavities, e.g.~a cubic one, this
condition cannot be fulfilled.
Thus the velocity effect does not contribute within the RWA 
(cf.~Ref.~\cite{dodonov}).

%%%%%%%%%%%%%%%%%%%%%%%%%%%%%%%%%%%%%%%%%%%%%%%%%%%%%%%%%%%%%%%%%%%%%%%%%%%%%%%
\subsection{Squeezing operator}\label{soperator}
%%%%%%%%%%%%%%%%%%%%%%%%%%%%%%%%%%%%%%%%%%%%%%%%%%%%%%%%%%%%%%%%%%%%%%%%%%%%%%%

In the following calculations we assume a case for which only the
squeezing term contributes (i.e.~the rather general case). 
The effective Hamiltonian can be derived 
immediately from the only contributing $\Delta\omega_I^2$-terms
\begin{eqnarray}
\int dt\,\hat H_1
&=&
\int dt\,\frac{\Delta\omega_I^2(t)}{4\omega_I}
\left(\hat a_I^\dagger e^{i\omega_I t}+
\hat a_I e^{-i\omega_I t}\right)^2
\nonumber\\
&=&
\frac{\omega_I\varepsilon}{2}
\int\limits_0^{\sf T} dt\,\sin(2\omega t)
\left(\hat a_I^\dagger e^{i\omega_I t}+
\hat a_I e^{-i\omega_I t}\right)^2
\nonumber\\
&\approx&
\frac{i\omega\varepsilon{\sf T}}{4}
\left((\hat a_1^\dagger )^2-
(\hat a_1)^2\right)
={\sf T}\,\hat H_1^{\rm eff}
\, .
\end{eqnarray}
Therefore the time evolution operator $\hat U$ coincides in the RWA
with a squeezing operator $\hat S_1$ for the lowest mode $I=1$
\begin{eqnarray}
\hat U
&=&
{\cal T}\left[\exp\left(-i\int dt\,\hat H_1(t)\right)\right]
\nonumber\\
&\approx&
\exp\left(\frac{\omega\varepsilon{\sf T}}{4}
\left((\hat a_1^\dagger )^2-
(\hat a_1)^2\right)\right)=\hat S_1
\,,
\end{eqnarray}
with a squeezing parameter $\Xi=\omega\varepsilon{\sf T}/2$.
This confirms the notion of the $\Delta\omega_I^2$-terms
in the perturbation Hamiltonian (\ref{cavH1}) as squeezing
contribution.
Having derived a closed expression for the approximate
time-evolution operator $\hat U\approx\hat S_1$ 
we are able to calculate the expectation 
value for the number operator to all orders in $\hat H_1^{\rm eff}$
within the RWA 
\begin{eqnarray}
\label{quetsch}
\left<\hat N_I\,\right>
&\approx&
{\rm Tr}\left\{
\hat\varrho_0\hat S_1^\dagger \hat N_I\hat S_1
\right\}=
\left<\hat N_I\,\right>_0
\nonumber\\
&&
+
\delta_{I1}
\sinh^2\left(\frac{\omega\varepsilon{\sf T}}{2}\right)
\left(1+2\left<\hat N_1\,\right>_0\right)
\, .
\end{eqnarray}
This non-perturbative result states that also at finite temperature
the number of photons $\Delta N_1$ created resonantly in the lowest
cavity mode increases exponentially.
The vacuum creation rate 
$\Delta N_1^{T=0}=\sinh^2(\omega\varepsilon{\sf T}/2)$
(see Ref.~\cite{dodonov}) gets enhanced by a thermal distribution
factor. 
Since the effective Hamiltonian becomes diagonal in the resonance case
such a behavior is consistent with the results in
Sec.~\ref{transformation}.  

%%%%%%%%%%%%%%%%%%%%%%%%%%%%%%%%%%%%%%%%%%%%%%%%%%%%%%%%%%%%%%%%%%%%%%%%%%%%%%%
\subsection{Local quantities}\label{local}
%%%%%%%%%%%%%%%%%%%%%%%%%%%%%%%%%%%%%%%%%%%%%%%%%%%%%%%%%%%%%%%%%%%%%%%%%%%%%%%

So far we have considered merely the expectation values of global 
observables such as particle number and energy.
But the canonical formalism developed here is also 
capable for investigating local quantities.
As an example we may consider the two-point function
\begin{eqnarray}
\langle\hat\Phi(\f{r})\hat\Phi(\f{r}'\,)\rangle=
{\rm Tr}\left\{\hat\varrho\,\hat\Phi(\f{r})\hat\Phi(\f{r}'\,)\right\}
\,.
\end{eqnarray}
According to the results of the previous Sections within the RWA the time 
evolution operator appears as a squeezing operator for the lowest mode
$I=1$. Expanding the field $\hat\Phi(\f{r})$ into the modes $f_I$ yields
\begin{eqnarray}
\langle\hat\Phi(\f{r})\hat\Phi(\f{r}'\,)\rangle=
\sum\limits_{IJ}{\rm Tr}
\left\{\hat\varrho_0\hat S_1^\dagger\hat Q_I\hat Q_J\hat S_1\right\}
f_I(\f{r})f_J(\f{r}'\,)
\,.
\end{eqnarray}
For $I \neq J$ the trace above vanishes and for $I=J\neq1$
it coincides with the undisturbed (thermal) expression.
Hence the only interesting case is $I=J=1$. In this situation 
the amplitudes $\hat Q_1$ are squeezed by the time-evolution operator 
$\hat S_1$. As a result the change of the correlation function induced by
the dynamics of the cavity can be cast into the form
\begin{eqnarray}
\Delta\langle\hat\Phi(\f{r})\hat\Phi(\f{r}'\,)\rangle
&=&
\left({e^{2\Xi}}-1\right){\rm Tr}\left\{\hat\varrho_0\hat Q_1^2\right\}
\nonumber\\
&&\times
f_1(\f{r})f_1(\f{r}'\,)
\nonumber\\
&=&
\left({e^{2\Xi}}-1\right)\left(1+2\left<\hat N_1\,\right>_0\right)
\nonumber\\
&&\times
f_1(\f{r})f_1(\f{r}'\,)/(2\omega_1)
\nonumber\\
&=&
\left({e^{2\Xi}}-1\right)
\left(
1+\frac{2}{\exp(\beta\omega_1)-1}
\right)
\nonumber\\
&&\times
f_1(\f{r})f_1(\f{r}'\,)/(2\omega_1)
\,,
\end{eqnarray}
where $\Xi$ again denotes the squeezing parameter.
In complete analogy one obtains the change of the correlation of the
field momenta
\begin{eqnarray}
\Delta\langle\hat\Pi(\f{r})\hat\Pi(\f{r}'\,)\rangle
&=&
\left({e^{-2\Xi}}-1\right)
\left(
1+\frac{2}{\exp(\beta\omega_1)-1}
\right)
\nonumber\\
&&\times
f_1(\f{r})f_1(\f{r}'\,)\omega_1/2
\,.
\end{eqnarray}
Note, that the canonical variables $\hat Q_I$ and the momenta 
$\hat P_I$ transform in an opposite way under squeezing: 
$\hat Q_I\rightarrow{e^{\Xi}}\hat Q_I$ whereas 
$\hat P_I\rightarrow{e^{-\Xi}}\hat P_I$.

These ingredients enable us to calculate the expectation values of the
energy-momentum tensor 
$T_{\mu\nu}=\partial_{\mu}\Phi\partial_{\nu}\Phi-
g_{\mu\nu}\partial_\rho\Phi\partial^\rho\Phi/2$.
The expression for the change of the energy density reads
\begin{eqnarray}
\Delta\langle\hat T_{00}\rangle
&=&
\nonumber
\frac{1}{4}
\left(
1+\frac{2}{\exp(\beta\omega_1)-1}
\right)
\\
&&
\times
\left[
\left({e^{2\Xi}}-1\right)(\nabla f_1)^2/\omega_1+
\left({e^{-2\Xi}}-1\right)f_1^2\omega_1
\right]
\,.
\nonumber\\
\end{eqnarray}
The stress tensor $T_{ij}$ consisting of purely spatial components
of $T_{\mu\nu}$ can be calculated in a completely analogue manner --
one would obtain additional terms like 
$\partial_i f_1 \partial_j f_1$. This quantity can be used to deduce
the mechanical properties, e.g.~the pressure, of the radiation field
inside the cavity.
The expectation value of the energy flux density $T_{0i}$ vanishes
within the RWA, since the change of the energy is always one power of
the vibration time $\sf T$ lower than the energy itself.

The above expression can be used to deduce the spatial distribution of
the  energy created by the dynamical perturbation of the cavity.
Since the squeezing parameter $\Xi$ is proportional to the vibration
time 
$\sf T$, the first term at the r.h.s.~dominates for large time
durations  
$\sf T$. In this situation the produced energy density behaves as 
$(\nabla f_1)^2$. For Dirichlet boundary conditions the 
{{\em eigen\,}}functions vanish 
at the boundary but their derivatives usually reach their maximum
value there. 
In the center of the cavity the lowest {{\em eigen\,}}function assumes
its maximum and -- consequently -- its derivative vanishes.
Ergo the energy density is concentrated near the boundaries of the
cavity in the case under consideration. 

However, this assertion crucially depends on the imposed boundary
conditions. 
For Neumann conditions the behavior is actually opposite.
In order to analyze the situation for the more realistic  
photon field one has to consider the corresponding 
{{\em eigen\,}}modes for this case. 
The modes of the electromagnetic field are more complicated and cannot 
be 
reduced to purely Dirichlet or Neumann conditions in general.
As demonstrated in Ref.~\cite{jackson}, one may divide the electromagnetic
{{\em eigen\,}}functions into TE and TM modes. E.g., for a cylinder with
perfectly  
conducting walls the energy in the lowest TM mode is concentrated near the
shell and vanishes at the symmetry axis, see \cite{jackson}. 
In addition, the {{\em eigen\,}}frequency of this particular mode does
not depend on the  
separation of the bottom and the top surface. Hence it cannot be excited 
resonantly by a vibration of the piston but only by a change of the radius.
In contrast, the lowest TE mode does depend on the separation and thus may 
be excited by a vibration of the piston.
In summary, the conclusions concerning the energy distribution of the 
photon field crucially depend on the particular mode that is excited by 
the cavity vibration.

%%%%%%%%%%%%%%%%%%%%%%%%%%%%%%%%%%%%%%%%%%%%%%%%%%%%%%%%%%%%%%%%%%%%%%%%%%%%%%%
\subsection{Discussion}\label{discussion-cavity}
%%%%%%%%%%%%%%%%%%%%%%%%%%%%%%%%%%%%%%%%%%%%%%%%%%%%%%%%%%%%%%%%%%%%%%%%%%%%%%%

In order to indicate the experimental relevance of the
calculations above one may specify the characteristic parameters.
For room temperature $\approx290\,\rm K$, which corresponds to thermal
wave lengths of about $50\,\mu\rm m$ and considering a cavity
of a typical size $\Lambda\approx1\,\rm cm$ one obtains a thermal
factor $(1+2\langle\hat N_1\,\rangle_0)={\cal O}[10^3]$.
As a consequence the number of photons $\Delta N_1$ created by the 
dynamical Casimir effect (after the vibration time $\sf T$) at the 
given temperature $T$ will be several orders of magnitude
larger in comparison with the pure vacuum effect at $T=0$, 
see \cite{letter}.   

This enhancement occurring at finite temperatures could be exploited
in experiments to verify the phenomenon of quantum radiation
as long as back-reaction processes (and losses, etc.) can be neglected.
Of course, one has also to take into account the number of
photons $\langle\hat N_I\,\rangle_0$ present at the temperature $T$
and their thermal variance  
\begin{eqnarray}
\sqrt{\sigma^2_0(N_I)}
&=&
\sqrt{\left<\hat N_I^2\right>_0-
\left<\hat N_I\,\right>^2_0}
=
\sqrt{\left<\hat N_I\,\right>_0+
\left<\hat N_I\,\right>^2_0}
\nonumber\\
&=&
\left<\hat N_I\,\right>_0
\left(1+{\cal O}\left[\frac{1}{N_I}\right]\right)
\,.
\end{eqnarray}
The latter quantity reflects the statistical uncertainty when
measuring the number of photons at a given temperature.
In order to obtain a number of created particles $\Delta N_1$
which is not much smaller than the corresponding thermal
variance $\sqrt{\sigma^2_0(N_1)}$ one has to ensure conditions that
will lead to a significant vacuum effect as well.
This implies that the argument of the hyperbolic sine function
in Eq.~(\ref{quetsch}), i.e.~the squeezing parameter
$\Xi=\omega\varepsilon{\sf T}/2$ should be at least of order one.
An estimate of the maximum value of the of the dimensionless amplitude
of the resonance wall vibration $\varepsilon_{\rm max}<10^{-8}$ is
given in Ref.~\cite{dodonov}.
For a characteristic size of the cavity of about $1\,\rm cm$ corresponding 
to a fundamental frequency $\omega\approx150\,\rm GHz$ the squeezing
parameter $\Xi$ approaches $1$ after several milliseconds.
It still remains as a challenge whether or not the requirement
$\Xi={\cal O}[1]$ could be achieved in a realistic experiment.
But -- provided an experimental device for generating a considerable
vacuum contribution becomes feasible -- there will be a strong
enhancement of the dynamical Casimir effect at finite temperatures.

%%%%%%%%%%%%%%%%%%%%%%%%%%%%%%%%%%%%%%%%%%%%%%%%%%%%%%%%%%%%%%%%%%%%%%%%%%%%%%%
%%%%%%%%%%%%%%%%%%%%%%%%%%%%%%%%%%%%%%%%%%%%%%%%%%%%%%%%%%%%%%%%%%%%%%%%%%%%%%%
\section{Dielectric media}\label{dielectric}
%%%%%%%%%%%%%%%%%%%%%%%%%%%%%%%%%%%%%%%%%%%%%%%%%%%%%%%%%%%%%%%%%%%%%%%%%%%%%%%
%%%%%%%%%%%%%%%%%%%%%%%%%%%%%%%%%%%%%%%%%%%%%%%%%%%%%%%%%%%%%%%%%%%%%%%%%%%%%%%

The previous Sections were devoted to the investigation of a scalar
field confined in a trembling cavity with Dirichlet boundary
conditions simulating perfect conductors.
Although the effect of quantum radiation has not yet been verified
conclusively in an according experiment, a resonantly vibrating cavity   
is expected to provide one of the most promising scenarios for this
aim.  
Of course, the assumption of perfectly conducting walls 
of the cavity is an idealization. One possible step towards a more
realistic description is to consider a dielectric medium with a finite 
permittivity $\epsilon$. 
Of course, one may also take into account the permeability, 
see e.g.~\cite{quant-rad}. But for reasons of simplicity we restrict
our further considerations to a purely dielectric medium.  

The quantum radiation generated by a moving body with a finite
refractive index was studied in Ref.~\cite{barton}, for example.
More generally, one may consider a medium with an arbitrary changing
permittivity $\epsilon(t,\f{r})$ and a local velocity field
$\f{v}(t,\f{r})$. 
Again these properties of the medium are treated classically, i.e.~as
an external background field. As the quantum field propagating in this 
background we consider the electromagnetic field. 
For non-relativistic velocities of the medium the Lagrangian density
governing the dynamics of the electromagnetic field is given by 
(see e.g.~\cite{quant-rad,eberlein})  
\begin{eqnarray}
\label{lag-diel}
{\cal L}=\frac{1}{2}\left(\epsilon\,\f{E}^2-\f{B}^2\right)+
(\epsilon-1)\,\f{v}\cdot\left(\f{E}\times\f{B}\right)
\,.
\end{eqnarray}
The particle definition for this vector field requires additional
considerations. Since it is described by a gauge invariant theory, it
possesses primary and secondary constraints, see
e.g.~\cite{teitelboim}. In Ref.~\cite{quant-rad} these gauge problems
are solved by virtue of the reduction of variables.      
However, other procedures, e.g.~the Dirac quantization, lead to the
same results \cite{alex}.

There have been various efforts to discover effects of quantum
radiation for such dynamical dielectrics: One interesting idea goes
back to Schwinger \cite{schwinger}, who suggested to explain the
phenomenon of sonoluminescence by this mechanism.
Sonoluminescence means the conversion of sound into light. 
In an according experiment one generates sound waves in a liquid
(water) in such a way that tiny oscillating bubbles emerge. 
Under appropriate conditions these bubbles emit light pulses, 
see e.g.~\cite{putterman} and references therein.
In spite of the considerable amount of work and the controversial
discussions in order to clarify the
relevance of quantum radiation with respect to sonoluminescence, see
e.g.~\cite{eberlein}, \cite{stefano}, \cite{gegensono} and also 
\cite{bordag1,bordag2}, there are still open questions,
since the dynamics of the bubble and the behavior 
in its interior are not known sufficiently.
We shall return to this point later on.

In view of the Lagrangian density in Eq.~(\ref{lag-diel}) the dynamical
properties of the medium enter in two terms: 
In analogy to the cavity example we may distinguish between the
squeezing effect due to a varying permittivity $\epsilon(t,\f{r})$
and the velocity effect governed by $\f{v}(t,\f{r})$.
In the following we consider situations where the squeezing term
gives the dominant contribution (see Ref.~\cite{quant-rad}, Section {\bf V})
and neglect the velocity field ($\f{v}=\f{0}\,$).
In contrast to the cavity example the squeezing term of the perturbation
Hamiltonian will not be diagonal in general
\begin{eqnarray}
\hat H_1(t)
&=&
\int d^3r\,\frac{1}{2}
\left(\frac{1}{\epsilon(t,\f{r})}-\frac{1}{\epsilon_\infty}\right)
\hat{\f{\Pi}}^2(t,\f{r})
\nonumber\\
&=&
\int d^3r\,\theta(t,\f{r})\,
\hat{\f{\Pi}}^2(t,\f{r})
\, .
\end{eqnarray}
$\hat{\f{\Pi}}=\hat{\f{E}}$ denotes the canonical momentum density
associated to the vector potential $\hat{\f{A}\;}\!$, see
e.g.~\cite{quant-rad} and \cite{diplom}.
$\theta(t,\f{r})$ symbolizes the deviation of the permittivity 
$\epsilon(t,\f{r})$ from its asymptotic value
$\epsilon_\infty=
\epsilon(|t|\uparrow\infty,\f{r})=
\epsilon(t,|\f{r}|\uparrow\infty)$.

The diagonalization of the undisturbed Hamiltonian via a particle
definition can be achieved with photons labeled by
$I=\{\f{k}\nu\}$ where $\f{k}$ denotes the wave vector of the 
photon and $\nu$ counts the two possible polarizations.
Within this basis the ${\cal S}$ and ${\cal U}$ matrices
assume the form
\begin{eqnarray}
\label{mediumS}
{\cal S}_{\fk{k}\nu,\fk{k}'\nu'}
&=&
-\frac{\sqrt{\omega_{\fk{k}}\omega_{\fk{k}'}}}{\mho}\,
({\f{e}_{\fk{k}\nu}}\cdot{\f{e}_{\fk{k}'\nu'}})
\nonumber\\
&&\times
\int d^4x\,\theta(\underline x)\,
\exp\left(i(\underline k+\underline k')\underline x\right) 
\nonumber\\
&=&
-\frac{\sqrt{\omega_{\fk{k}}\omega_{\fk{k}'}}}{\mho}\,
({\f{e}_{\fk{k}\nu}}\cdot{\f{e}_{\fk{k}'\nu'}})\;
\widetilde\theta(\underline k+\underline k')
\,,
\end{eqnarray}
where $\mho$ denotes the quantization volume, and 
\begin{eqnarray}
\label{mediumU}
{\cal U}_{\fk{k}\nu,\fk{k}'\nu'}
=
\frac{\sqrt{\omega_{\fk{k}}\omega_{\fk{k}'}}}{\mho}\,
({\f{e}_{\fk{k}\nu}}\cdot{\f{e}_{\fk{k}'\nu'}})\;
\widetilde\theta(\underline k-\underline k')
\, ,
\end{eqnarray}
respectively.
In complete analogy to the cavity we may calculate of the quadratic
response of the number operator 
\begin{eqnarray}
\label{mediumN}
\Delta N_{\fk{k}\nu}
&=&
\sum\limits_{\fk{k}'\nu'}
\left|{\cal S}_{\fk{k}\nu,\fk{k}'\nu'}\right|^2
\left(1+\left<\hat N_{\fk{k}'\nu'}\right>_0
+\left<\hat N_{\fk{k}\nu}\right>_0\right)
\nonumber\\
&&+
\sum\limits_{\fk{k}'\nu'}
\left|{\cal U}_{\fk{k}\nu,\fk{k}'\nu'}\right|^2
\left(\left<\hat N_{\fk{k}'\nu'}\right>_0
-\left<\hat N_{\fk{k}\nu}\right>_0\right)
\,.
\end{eqnarray}
It is again possible to recognize the thermal corrections
to the vacuum effect $\sum\limits_{\fk{k}'\nu'}
\left|{\cal S}_{\fk{k}\nu,\fk{k}'\nu'}\right|^2$.
In Ref.~\cite{quant-rad} we gave a general proof that for
massless and non-self-interacting bosonic fields at zero
temperature the spectral energy density $e(\omega)$ created
by smooth and localized disturbances behaves as
$e(\omega)\sim\omega^4$ for small $\omega$.
As one can easily check in the equation above this is no
longer valid in general at finite temperatures due to the
Boltzmann distribution function that becomes singular with
$1/\omega$ for small $\omega$.

%%%%%%%%%%%%%%%%%%%%%%%%%%%%%%%%%%%%%%%%%%%%%%%%%%%%%%%%%%%%%%%%%%%%%%%%%%%%%%%
\subsection{Small $R$ expansion}\label{smallR}
%%%%%%%%%%%%%%%%%%%%%%%%%%%%%%%%%%%%%%%%%%%%%%%%%%%%%%%%%%%%%%%%%%%%%%%%%%%%%%%

The structure of Eq.~(\ref{mediumN}) is too complicated for a general 
discussion of the physical properties of the induced quantum radiation
by means of simple expressions.
For that purpose it is necessary to use some approximations.
One possibility is to assume that the region where the permittivity
$\epsilon(t,\f{r})$ differs from its asymptotic value
$\epsilon_\infty=
\epsilon(|t|\uparrow\infty,\f{r})=
\epsilon(t,|\f{r}|\uparrow\infty)$ is very small.
This assumption can be used to expand the Fourier transform of the
perturbation function $\theta(t,\f{r})$ in powers of $R$, where $R$
denotes a characteristic length scale of the disturbance
\begin{eqnarray}
\widetilde\theta(\underline k)
&=&
\int dt\, e^{i\omega t}\int d^3r\,e^{i\fk{k}\cdot\fk{r}}
\theta(t,\f{r})
\nonumber\\
&=&
\theta_0\int dt\, e^{i\omega t}\,{V}(t)+{\cal O}[R^4]
\nonumber\\
&=&
\theta_0\widetilde{V}(\omega)+{\cal O}[R^4]
\, .
\end{eqnarray}
For the lowest (volume ${V}\sim R^3$) term of this expansion it is
possible to calculate the associated radiated energy in close analogy
to the moving mirror example.
But in the case under consideration the evaluations have to be accomplished 
in 3+1 dimensions which leads to additional scale factors $k^2$ and $k'^2$ 
due to the $d^3k$- and $d^3k'$-integrations. 
This results in the occurrence of Hurwitz zeta-functions of higher order 
$\zeta(4,1-i(t-t')/\beta)$ and $\zeta(5,1-i(t-t')/\beta)$ and therefore
in higher powers of the temperature $T$.
After some calculations the lowest order terms in $R$ and $T$ of the
total radiated energy yield
\begin{eqnarray}
E
&=&
\left(\frac{\epsilon_\infty}{2\pi}\right)^3
\frac{\theta_0^2}{3\cdot5\cdot7}
\nonumber\\
&&\times
\left(
\int dt\,{\stackrel{....}{V}\,}^2+
\,\zeta(4)\,\frac{8!}{2!\,4!}\,
T^4\int dt\,\ddot{V}^2
\right)
\,.
\end{eqnarray}
In analogy to the moving mirror example in Sec.~\ref{moving} the first
term describes the (lowest) vacuum contribution, which was already
obtained in Ref.~\cite{quant-rad}, whereas the second term represents
the (lowest) temperature correction.
But in contrast to the moving mirror in 1+1 dimensions the
lowest thermal correction increases with $T^4$ in this scenario owing to
the additional $k,k'$-integrations.
 
If we would have the exact data for the oscillating bubble we were able
to evaluate the number of photons generated by the quantum radiation
and so we could quantify the contribution of this mechanism to 
the phenomenon of sonoluminescence -- under the assumptions made. 
But as these data are not known yet with a sufficient accuracy, 
this question remains unsolved at this stage.

%%%%%%%%%%%%%%%%%%%%%%%%%%%%%%%%%%%%%%%%%%%%%%%%%%%%%%%%%%%%%%%%%%%%%%%%%%%%%%%
\subsection{Large $R$ limit}\label{largeR}
%%%%%%%%%%%%%%%%%%%%%%%%%%%%%%%%%%%%%%%%%%%%%%%%%%%%%%%%%%%%%%%%%%%%%%%%%%%%%%%

Now we consider the opposite situation and assume that
the permittivity changes over very large volumes in the same way. 
In such a scenario the disturbance function $\theta$ becomes nearly
position-independent $\theta(t,\f{r})\approx\theta(t)$.
In this limit it is also possible to simplify Eq.~(\ref{mediumN})
since the $d^3r$-integrations in Eqs.~(\ref{mediumS}) and 
(\ref{mediumU}) produce $\delta^3(\f{k}\pm\f{k}')$-distributions 
and therefore the mode integrations break down.   
As a result the expression for the quadratic response of the
number operator obeys a structure similar to the squeezing
term in the cavity example
\begin{eqnarray}
\Delta N_{\fk{k}\nu}
=
\left|{\cal S}_{\fk{k}\nu,(-\fk{k})\nu}\right|^2
\left(1+2\left<\hat N_{\fk{k}\nu}\right>_0\right)
\, .
\end{eqnarray}
To establish the analogy once more we note that a 
perturbation like $\theta(t,\f{r})=\varepsilon\sin(2\omega t)$
generates also a generalized squeezing operator 
\begin{eqnarray}
\hat U(\Xi)=\exp\left(\;\frac{\Xi}{2}\;
\sum\limits_{\nu,|\fk{k}|=\omega}
\left(\hat A_{\fk{k}\nu}^\dagger\hat A_{(-\fk{k})\nu}^\dagger-
{\rm h.c.}
\right)
\right)
\,,
\end{eqnarray}
similar to the resonantly vibrating cavity, see also \cite{birula}. 
However, there is a crucial difference between the medium and the cavity:
In a closed cavity there exist only particles with special discrete
frequencies ({{\em eigen\,}}values of the Laplace operator). In contrast,
for the dielectric medium without boundary conditions all 
positive frequencies are occupied by photons.  
Hence one has to vibrate a (finite) cavity with a special 
(resonance) frequency in order to create particles while in a medium
the frequency may be arbitrary. 

Investigating the two-photon correlation 
${\cal C}_{\fk{k}\nu,\fk{k}'\nu'}$
at zero temperature one observes that in this case the photons 
are most probably emitted back-to-back:
${\cal C}_{\fk{k}\nu,\fk{k}'\nu'}(T=0)\sim
\delta(\f{k}+\f{k}')\delta_{\nu\nu'}$. 
At finite temperatures the second term in Eq.~(\ref{correl}) gives
raise to an additional negative correlation which is isotropic,
i.e.~it does not depend on the directions of propagation of the
two photons.     
However, the anisotropic and temperature-independent back-to-back 
correlation represents one possibility to distinguish between the
photons arising from the quantum radiation and the purely thermal
radiation. 
This observation (see \cite{stefano}) might perhaps be used to
clarify the origin of the photons (i.e.~the underlying mechanism)
within the phenomenon of sonoluminescence.  
 
%%%%%%%%%%%%%%%%%%%%%%%%%%%%%%%%%%%%%%%%%%%%%%%%%%%%%%%%%%%%%%%%%%%%%%%%%%%%%%%
%%%%%%%%%%%%%%%%%%%%%%%%%%%%%%%%%%%%%%%%%%%%%%%%%%%%%%%%%%%%%%%%%%%%%%%%%%%%%%%
\section{Friedmann-Robertson-Walker metric}\label{robertson}
%%%%%%%%%%%%%%%%%%%%%%%%%%%%%%%%%%%%%%%%%%%%%%%%%%%%%%%%%%%%%%%%%%%%%%%%%%%%%%%
%%%%%%%%%%%%%%%%%%%%%%%%%%%%%%%%%%%%%%%%%%%%%%%%%%%%%%%%%%%%%%%%%%%%%%%%%%%%%%%

In the previous Sections we focused our attention on mirrors
represented by Dirichlet boundary conditions and on dielectric media. 
Now we are going to apply the canonical formalism developed there to
yet another scenario -- where the gravitational field generates
quantum radiation. 
  
According to the commonly suggested scenario the cosmological evolution
starts at a stage of high temperatures. It is generally believed that
the back-reaction of the cosmological particle production onto the
gravitational sector yields a significant contribution.
Consequently, it will be important to calculate the temperature
effects that could affect the cosmological dynamics at very early
stages.

Let us consider the minimally coupled massless scalar field
propagating in the conformally flat Friedmann-Robertson-Walker
space-time, see e.g.~Ref.~\cite{audretsch1} for a related calculation 
at zero temperature.

The Friedmann-Robertson-Walker metric represents a solution of
Einstein's equations for a homogeneous and isotropic distribution of
matter and describes an expanding (or contracting) universe. 
Depending on the density $\varrho$ of the matter   
(if we omit the cosmological constant) there exist three different
branches of this solution: 
For densities $\varrho$ exceeding the critical density 
$\varrho_{\rm c}$ one obtains the closed elliptical universe, which
eventually re-collapses. 
For $\varrho<\varrho_{\rm c}$ one is lead to the open hyperbolic
universe and for $\varrho=\varrho_{\rm c}$ we get the open   
conformally flat Friedmann-Robertson-Walker space-time.
In contrast to the first case ($\varrho>\varrho_{\rm c}$) the other
scenarios ($\varrho\leq\varrho_{\rm c}$) imply an eternally expanding
universe. 

In order to specify the correct value of the density $\varrho$ one has
to deal with the problem of the unknown dark matter.
In view of the present status of the observations it might well be
possible that the density $\varrho$ is indeed close or equal to the
critical value $\varrho_{\rm c}$, 
which is connected to the Hubble constant. 
In any case, for small space-time domains the conformally flat
Friedmann-Robertson-Walker space-time should be a good approximation
in any case. 
In terms of the conformal coordinates $(t,\f{r})$ the corresponding
metric is given by
\begin{eqnarray}
ds^2=\Omega^2(t)\left(dt^2-d\f{r}^2\right)
\, ,
\end{eqnarray}
where $\Omega^2(t)$ denotes the scale factor governing the Hubble
expansion. It describes the change of the measure of length and time
scales during the cosmological evolution, 
e.g.~inducing the cosmological red-shift.  
  
However, the following calculations will become easier if we introduce
a slightly different time coordinate $t\rightarrow\tau$ with
\begin{eqnarray}
d\tau=\Omega^{-2}dt
\,,
\end{eqnarray}
see also Ref.~\cite{audretsch1}.
In terms of the new time coordinate the metric can be cast into 
the form 
\begin{eqnarray}
ds^2=\Omega^6(\tau)d\tau^2-\Omega^2(\tau)d\f{r}^2
\, .
\end{eqnarray}
The action for a minimally coupled massless scalar field 
propagating in this particular curved space-time reads
\begin{eqnarray}
{\cal A}
&=&
\frac{1}{2} \int d^4x\,\sqrt{-g}\,
\partial_\mu\Phi\,g^{\mu\nu}\,\partial_\nu\Phi
\nonumber\\
&=&
\frac{1}{2}
\int d\tau\int d^3r\,\left(\dot\Phi^2-\Omega^4(\nabla\Phi)^2\right)
\, .
\end{eqnarray}
As the advantage of the new time coordinate $\tau$ we observe the
cancellation of the scale factor in front of the $\dot\Phi^2$-term.
Consequently the equation of motion assumes the simple form 
\begin{eqnarray}
\frac{\partial^2\Phi}{\partial\tau^2}
=\Omega^4(\tau)\nabla^2\Phi
\, .
\end{eqnarray}
After the usual canonical quantization procedure the Hamiltonian can
be cast into the form  
\begin{eqnarray}
\hat H(\tau)
=
\frac{1}{2}
\int d^3r\left(\hat\Pi^2+\Omega^4(\tau)(\nabla\hat\Phi)^2\right)
\,.
\end{eqnarray}
One observes a close similarity to the large $R$ limit in Section 
\ref{largeR}.
As we shall see below, this similarity holds also for the number of 
created particles. 

In complete analogy to Sec.~\ref{trembling} it is possible to diagonalize
this time-dependent Hamiltonian by an expansion of the field
$\hat\Phi$ into a complete set of orthogonal {{\em eigen\,}}functions
of the Laplace operator. 
Owing to the spatial homogeneity and isotropy of the space-time there
is an ambiguity concerning the selection of such a basis set. 
Here we choose the {{\em eigen\,}}functions to be completely
time-independent, $f_{I}=f_{I}(\f{r})$.
As a consequence, adopting this expansion  
\begin{eqnarray}
\hat\Phi(t,\f{r})=\sum\limits_{{I}}\hat Q_{I}(t)f_{I}(\f{r})
\,,
\end{eqnarray}
the resulting modes $\hat Q_{I}(t)$ do not obey any
inter-mode interaction due to the spatial integration and
the orthogonality and time-independence of the {{\em eigen\,}}functions 
$f_{I}$.      
Hence the time-dependent Hamiltonian is diagonal
\begin{eqnarray}
\hat H(\tau)=
\frac12
\sum\limits_{{I}}
\left(
\hat P^2_{I}(\tau)
+\Omega^4(\tau)\,\omega_{I}^2\,\hat Q_{I}^2(\tau)
\right)
\,,
\end{eqnarray}
where $-\omega_{I}^2$ denote the time-independent 
{{\em eigen\,}}values of the Laplacian.
Now we may use the outcome of Section \ref{transformation} and
we arrive at  
\begin{eqnarray}
\Delta N_\omega^{T}=
\Delta N_\omega^{T=0}
\left(1+\frac{2}{\exp(\beta\omega)-1}\right)
\, .
\end{eqnarray}
It should be mentioned here that the particle number above is --
strictly speaking -- merely a formal quantity since it does not
describe particles in a well-defined and unique sense.
The Friedmann-Robertson-Walker space-time is not time-translationally
invariant and thus does not possess a corresponding Killing-vector.
Ergo the definition of energy necessitates additional considerations.
It is not possible to define a physical reasonable {\em and}
conserved energy. Of course, this fact is consistent with the permanent
particle creation. 
Hence the interpretation of the above quantity $\Delta N_\omega$ is
not obvious -- at zero as well as at finite temperatures, see also
Ref.~\cite{audretsch2}. 
But here we are mainly interested in the influence of finite
(initial) temperatures.   
Fortunately, the finite temperature effects factorize out and the
problems mentioned above concern the pre-factor 
$\Delta N_\omega^{T=0}$ only.
In summary we may draw the conclusion that -- putting aside the
problem of the interpretation of the vacuum term 
$\Delta N_\omega^{T=0}$ -- the 
particle creation in the Friedmann-Robertson-Walker space-time at
finite (initial) temperatures gets strongly enhanced by a thermal
factor in analogy to the resonantly vibrating cavity.      

It should be mentioned here that the phenomenon of particle creation 
(quantum radiation) in the conformally flat Friedmann-Robertson-Walker 
space-time can be observed merely for fields which are not conformally
invariant.  
As counter-examples we may quote the massless scalar field in 1+1
dimensions, the massless and conformally coupled scalar field in 3+1
dimensions (see also Ref.~\cite{audretsch1}), and -- last but not
least -- the electromagnetic field in 3+1 dimensions.
Obviously the absence of any mass terms is essential for the conformal
invariance. 
For these conformally invariant fields the equation of motion does not
lead to any mixing of positive and negative frequency solutions within
the conformally flat Friedmann-Robertson-Walker metric and thus no
particles are created.   

%%%%%%%%%%%%%%%%%%%%%%%%%%%%%%%%%%%%%%%%%%%%%%%%%%%%%%%%%%%%%%%%%%%%%%%%%%%%%%%
%%%%%%%%%%%%%%%%%%%%%%%%%%%%%%%%%%%%%%%%%%%%%%%%%%%%%%%%%%%%%%%%%%%%%%%%%%%%%%%
\section{Summary}
%%%%%%%%%%%%%%%%%%%%%%%%%%%%%%%%%%%%%%%%%%%%%%%%%%%%%%%%%%%%%%%%%%%%%%%%%%%%%%%
%%%%%%%%%%%%%%%%%%%%%%%%%%%%%%%%%%%%%%%%%%%%%%%%%%%%%%%%%%%%%%%%%%%%%%%%%%%%%%%

Calculating the number of particles created by dynamical external
conditions we found that for the case of a completely diagonal
Hamiltonian the number of produced particle at finite temperature
equals the analogue quantity at zero temperature times a thermal
factor.   

Focusing on the scenario of a resonantly vibrating cavity we were able
to derive  within the RWA the effective perturbation Hamiltonian which
turned out to be diagonal. As a consequence we observe an enhancement
of the dynamical Casimir effect at finite temperatures as described
above.  

In contrast to this non-perturbative result the finite temperature
corrections to the energy radiated by a single moving mirror in 1+1
dimensions was calculated within response theory. 

In close analogy we derived the energy of the photons generated by an
bubble with an oscillating radius within a dielectric medium.

Finally we investigated the particle production within the
Friedmann-Robertson-Walker universe at finite temperatures --
where the Hamiltonian can again cast into a diagonal form.

%%%%%%%%%%%%%%%%%%%%%%%%%%%%%%%%%%%%%%%%%%%%%%%%%%%%%%%%%%%%%%%%%%%%%%%%%%%%%%%
%%%%%%%%%%%%%%%%%%%%%%%%%%%%%%%%%%%%%%%%%%%%%%%%%%%%%%%%%%%%%%%%%%%%%%%%%%%%%%%
\section{Conclusion}
%%%%%%%%%%%%%%%%%%%%%%%%%%%%%%%%%%%%%%%%%%%%%%%%%%%%%%%%%%%%%%%%%%%%%%%%%%%%%%%
%%%%%%%%%%%%%%%%%%%%%%%%%%%%%%%%%%%%%%%%%%%%%%%%%%%%%%%%%%%%%%%%%%%%%%%%%%%%%%%

As a main result of this article we have presented a theoretical
description of quantum radiation at finite temperatures by
generalizing the canonical approach developed earlier 
\cite{trembling,diplom,quant-rad,blackholes,onhawking}. 
The major advantage of this formalism is its generality and flexibility.

Depending on the characteristic scales associated to the perturbation
the effects of finite temperatures represent potentially significant
contributions to the quantum radiation and hence should be taken into
account for realistic estimations of this striking effect. 

This observation may be interpreted in the following way: Not only the
vacuum fluctuations but also the thermal excitations are converted
into real particles by the influence of the dynamical external
conditions.

%%%%%%%%%%%%%%%%%%%%%%%%%%%%%%%%%%%%%%%%%%%%%%%%%%%%%%%%%%%%%%%%%%%%%%%%%%%%%%%
%%%%%%%%%%%%%%%%%%%%%%%%%%%%%%%%%%%%%%%%%%%%%%%%%%%%%%%%%%%%%%%%%%%%%%%%%%%%%%%
\section{Discussion}
%%%%%%%%%%%%%%%%%%%%%%%%%%%%%%%%%%%%%%%%%%%%%%%%%%%%%%%%%%%%%%%%%%%%%%%%%%%%%%%
%%%%%%%%%%%%%%%%%%%%%%%%%%%%%%%%%%%%%%%%%%%%%%%%%%%%%%%%%%%%%%%%%%%%%%%%%%%%%%%

For the perhaps most interesting scenario in view of a possible
experimental verification of the dynamical Casimir effect -- the
resonantly vibrating cavity -- we specified some relevant parameters
in Sec.~\ref{discussion-cavity}. 

But as it became evident in Section \ref{dielectric}, a cavity filled
with a dielectric medium with an resonantly oscillating permittivity  
$\epsilon(t)=\epsilon_\infty+\Delta\epsilon\sin(2 \omega t)$ generates
quite similar effects.
Identifying the dimensionless amplitude $\varepsilon$ of the vibration 
of the cavity in the first case with the relative change of the
permittivity $\Delta\epsilon$ in the second case the set of relevant
parameters is completely equivalent in both situations.

It turns out that the thermal factor enhancing the dynamical Casimir
effect is of order $10^3$. 
But this enormous amplification should be contrasted to the thermal
variance of the number of particles present initially which may
complicate the measurement of the number of produced particles.
For the relevant temperature regions both terms are of the same order
of magnitude. 
 
As a consequence the finite temperature effects do not necessarily
generate difficulties concerning the experimental verification of the
dynamical Casimir effect. 
From our point of view there is no need to perform an experiment at
low temperatures -- which might be much more involved than one at room
temperature. 

%%%%%%%%%%%%%%%%%%%%%%%%%%%%%%%%%%%%%%%%%%%%%%%%%%%%%%%%%%%%%%%%%%%%%%%%%%%%%%%
%%%%%%%%%%%%%%%%%%%%%%%%%%%%%%%%%%%%%%%%%%%%%%%%%%%%%%%%%%%%%%%%%%%%%%%%%%%%%%%
\section{Outlook}
%%%%%%%%%%%%%%%%%%%%%%%%%%%%%%%%%%%%%%%%%%%%%%%%%%%%%%%%%%%%%%%%%%%%%%%%%%%%%%%
%%%%%%%%%%%%%%%%%%%%%%%%%%%%%%%%%%%%%%%%%%%%%%%%%%%%%%%%%%%%%%%%%%%%%%%%%%%%%%%

There are several possibilities to extend the presented calculations
to more general scenarios:
Firstly, one may drop the idealization of a perfectly conduction
cavity and take losses into account \cite{gernot}.
In analogy, it would be interesting to study the effects of a
non-trivial dispersion relation of a dielectric medium in view of the
phenomenon of quantum radiation.
This might be especially interesting for the investigation of the
dielectric black hole analogues, see e.g.~\cite{DBHA}
The investigation of the phenomenon of quantum radiation for
interacting fields obeying non-linear equations of motion is rather
challenging. 

%%%%%%%%%%%%%%%%%%%%%%%%%%%%%%%%%%%%%%%%%%%%%%%%%%%%%%%%%%%%%%%%%%%%%%%%%%%%%%%
%%%%%%%%%%%%%%%%%%%%%%%%%%%%%%%%%%%%%%%%%%%%%%%%%%%%%%%%%%%%%%%%%%%%%%%%%%%%%%%
\section{Acknowledgments}
%%%%%%%%%%%%%%%%%%%%%%%%%%%%%%%%%%%%%%%%%%%%%%%%%%%%%%%%%%%%%%%%%%%%%%%%%%%%%%%
%%%%%%%%%%%%%%%%%%%%%%%%%%%%%%%%%%%%%%%%%%%%%%%%%%%%%%%%%%%%%%%%%%%%%%%%%%%%%%%

We thank A.~Calogeracos, V.~V.~Dodonov, and G.~Schaller
for fruitful discussions.
R.~S.~and G.~P.~appreciated the kind hospitality during their stays at  
the Atomic Physics Group of the Chalmers University in
Gothenburg/Sweden as well as at the Institute for Theoretical Physics of
the Debrecen University in Hungary. 
This stay was supported by DAAD and M\"OB.
R.~S.~is thankful to the Particle Physics Group at the
Kyoto University in Japan and to MONBUSHO. 
Financial support from BMBF, DFG, and GSI is gratefully acknowledged. 

%%%%%%%%%%%%%%%%%%%%%%%%%%%%%%%%%%%%%%%%%%%%%%%%%%%%%%%%%%%%%%%%%%%%%%%%%%%%%%%
%%%%%%%%%%%%%%%%%%%%%%%%%%%%%%%%%%%%%%%%%%%%%%%%%%%%%%%%%%%%%%%%%%%%%%%%%%%%%%%
\addcontentsline{toc}{section}{References}%%%%%%%%%%%%%%%%%%%%%%%%%%%%%%%%
%%%%%%%%%%%%%%%%%%%%%%%%%%%%%%%%%%%%%%%%%%%%%%%%%%%%%%%%%%%%%%%%%%%%%%%%%%%%%%%
%%%%%%%%%%%%%%%%%%%%%%%%%%%%%%%%%%%%%%%%%%%%%%%%%%%%%%%%%%%%%%%%%%%%%%%%%%%%%%%

\end{document}